\documentclass[%
    reprint,
    superscriptaddress,
    amsmath,amssymb,
    aps,
    prx,
    floatfix,
]{revtex4-2}

\usepackage[normalem]{ulem}

\usepackage{graphicx}
\usepackage{dcolumn}
\usepackage{bm}
\usepackage[binary-units=true]{siunitx}
\usepackage{upgreek}
\usepackage{color}
\usepackage{braket}
\usepackage{natbib}
\usepackage{booktabs}
\usepackage{multirow}
\usepackage[ttdefault=true]{AnonymousPro}

\usepackage{float}
\usepackage{capt-of}

\DeclareSIUnit{\sample}{Sa}

\definecolor{Q1}{RGB}{0,148,255}
\definecolor{Q2}{RGB}{255,106,0}
\definecolor{Q3}{RGB}{0,127,127}
\definecolor{Q4}{RGB}{255,0,0}
\definecolor{Q5}{RGB}{87,0,127}

\newcommand{\ReMo}[1]{{\color[rgb]{0,0,0}#1}}

\begin{document}

\title{Deep Neural Network Discrimination of Multiplexed Superconducting Qubit States}

\author{Benjamin Lienhard}
    \email{blienhar@mit.edu}
    \affiliation{Department of Electrical Engineering and Computer Science, Massachusetts Institute of Technology, Cambridge, MA 02139, USA}
    \affiliation{Research Laboratory of Electronics, Massachusetts Institute of Technology, Cambridge, MA 02139, USA}
\author{Antti Veps\"al\"ainen}
    \affiliation{Research Laboratory of Electronics, Massachusetts Institute of Technology, Cambridge, MA 02139, USA}
\author{Luke C. G. Govia}
    \email{luke.c.govia@raytheon.com}
    \affiliation{Quantum Engineering and Computing Group, Raytheon BBN Technologies, Cambridge, MA 02138, USA}
\author{Cole R. Hoffer}
    \affiliation{Department of Electrical Engineering and Computer Science, Massachusetts Institute of Technology, Cambridge, MA 02139, USA}
    \affiliation{Research Laboratory of Electronics, Massachusetts Institute of Technology, Cambridge, MA 02139, USA}
\author{Jack Y. Qiu}
    \affiliation{Department of Electrical Engineering and Computer Science, Massachusetts Institute of Technology, Cambridge, MA 02139, USA}
    \affiliation{Research Laboratory of Electronics, Massachusetts Institute of Technology, Cambridge, MA 02139, USA}
\author{Diego Rist\`e}
     \affiliation{Quantum Engineering and Computing Group, Raytheon BBN Technologies, Cambridge, MA 02138, USA}
\author{Matthew Ware}
     \affiliation{Quantum Engineering and Computing Group, Raytheon BBN Technologies, Cambridge, MA 02138, USA}
\author{David Kim}
    \affiliation{MIT Lincoln Laboratory, Lexington, MA 02421, USA}
\author{Roni Winik}
    \affiliation{Research Laboratory of Electronics, Massachusetts Institute of Technology, Cambridge, MA 02139, USA}
\author{Alexander Melville}
    \affiliation{MIT Lincoln Laboratory, Lexington, MA 02421, USA}
\author{Bethany Niedzielski}
    \affiliation{MIT Lincoln Laboratory, Lexington, MA 02421, USA}
\author{Jonilyn Yoder}
    \affiliation{MIT Lincoln Laboratory, Lexington, MA 02421, USA}
\author{Guilhem J. Ribeill}
     \affiliation{Quantum Engineering and Computing Group, Raytheon BBN Technologies, Cambridge, MA 02138, USA}
\author{Thomas A. Ohki}
     \affiliation{Quantum Engineering and Computing Group, Raytheon BBN Technologies, Cambridge, MA 02138, USA}
\author{Hari K. Krovi}
     \affiliation{Quantum Engineering and Computing Group, Raytheon BBN Technologies, Cambridge, MA 02138, USA}
\author{Terry P. Orlando}
    \affiliation{Department of Electrical Engineering and Computer Science, Massachusetts Institute of Technology, Cambridge, MA 02139, USA}
    \affiliation{Research Laboratory of Electronics, Massachusetts Institute of Technology, Cambridge, MA 02139, USA}
\author{Simon Gustavsson}
    \affiliation{Research Laboratory of Electronics, Massachusetts Institute of Technology, Cambridge, MA 02139, USA}
\author{William D. Oliver}
    \affiliation{Department of Electrical Engineering and Computer Science, Massachusetts Institute of Technology, Cambridge, MA 02139, USA}
    \affiliation{Research Laboratory of Electronics, Massachusetts Institute of Technology, Cambridge, MA 02139, USA}
    \affiliation{MIT Lincoln Laboratory, Lexington, MA 02421, USA}

\date{\today}

\begin{abstract}

Demonstrating \ReMo{a} quantum computational advantage will require high-fidelity control and readout of multi-qubit systems. As system size increases, multiplexed qubit readout becomes a practical necessity to limit the growth of resource overhead. Many contemporary qubit-state discriminators presume single-qubit operating conditions or require considerable computational effort, limiting their potential extensibility. Here, we present multi-qubit readout using neural networks as state discriminators. We compare our approach to contemporary methods employed on a quantum device with five superconducting qubits and frequency-multiplexed readout. We find that fully-connected feedforward neural networks increase the qubit-state-assignment fidelity for our system. Relative to contemporary discriminators, the assignment error rate is reduced by up to \SI{25}{\percent} due to the compensation of system-dependent nonidealities such as readout crosstalk which is reduced by up to one order of magnitude. Our work demonstrates a potentially extensible building block for high-fidelity readout relevant to both near-term devices and future fault-tolerant systems. 


\end{abstract}

\pacs{Valid PACS appear here}
\maketitle

\section{\label{sec:INT}Introduction}

Quantum computers hold the promise to solve particular computational tasks substantially faster than conventional computers~\cite{Grover1996_algo, Shor1996_algo}. Depending on the computational task, such quantum devices need to be composed of hundreds to millions of high-fidelity qubits. 
An increase from a few to many qubits is generally accompanied by the challenge of maintaining low error rates for qubit control and readout.

Over the past two decades, superconducting qubits have emerged as a leading quantum computing platform~\cite{Nori2017_review, Krantz2019_review}. Today, individual qubits with coherence times exceeding \SI{100}{\micro\second}~\cite{Jin2015_therm}, gate times of a few tens of nanoseconds~\cite{Arute2019_supremacy}, and \ReMo{individual} single- and two-qubit gate operation fidelities above the most lenient thresholds for quantum error correction have been demonstrated for devices with up to 50 qubits~\cite{Kjaergaard2019_review, Arute2019_supremacy}. However, considerable work is still needed to retain and even further improve these fidelities as systems increase in size and complexity~\cite{Gambetta2017_qc}. 

Errors arise during all stages of the circuit model: initialization~\cite{Riste2012_init, Johnson2012_init}, computation~\cite{Reed2012_QEC, Barends2014_QEC}, and readout~\cite{Krantz2016}. In many implementations, qubit readout plays a key role beyond merely measuring the computational output. For example, quantum error correction protocols require repeated readout of syndrome qubits~\cite{DiVincenzo2009_fault, Gambetta2017_qc, Fowler2012_QEC_SC}. Even without error correction, many of the noisy intermediate-scale quantum (NISQ)~\cite{Preskill2018_supremacy} era algorithms involve an iterative optimization that generates a target quantum state based on prior trial-state measurements of qubits~\cite{Peruzzo2014_algo_VQE,Farhi2014_algo_QAOA}. In addition, diagnosing qubit-readout errors in post-processing requires computationally expensive statistical analyses of repeated computation and measurement~\cite{Maciejewski2020_Rerr,Magesan2015_SVM, Arute2019_supremacy}. Developing accurate and resource-efficient qubit-state readout is a key to realize useful quantum information processing tasks. 

\begin{figure*}[t]
	\centering
	\includegraphics[width=0.94\textwidth]{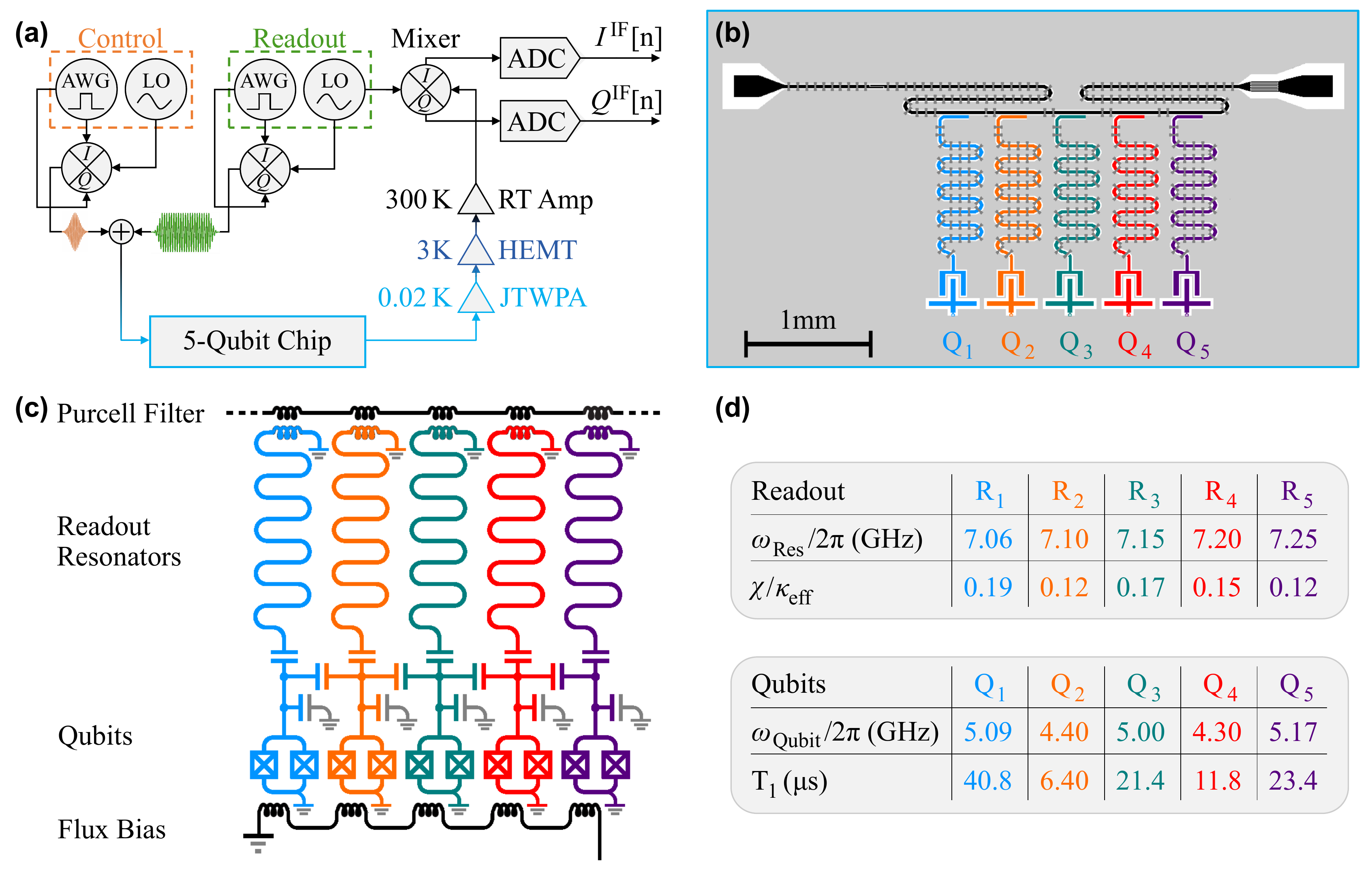}
	\caption{\label{fig:SCQC}Measurement Setup and Chip. (a) Schematic of superconducting qubit control and readout. The control and readout pulses, generated by an arbitrary waveform generator (AWG) and up-converted to \si{\giga\hertz} frequencies using a local oscillator (LO), are sent through attenuated signal lines to the readout resonator on the five-qubit chip. The transmitted readout signal is amplified by a Josephson traveling-wave parametric amplifier (JTWPA), a high-electron-mobility transistor (HEMT), and a room-temperature amplifier. Subsequently, the signal is down-converted to \si{\mega\hertz} frequencies and digitized---in-phase \(I^{\rm IF}[n]\) and quadrature \(Q^{\rm IF}[n]\) sequences at intermediate frequencies (IF). Colored optical micrograph (b) and the circuit schematic (c) comprising five superconducting transmon qubits. The qubit transition frequencies are tuned via a global flux bias. Each qubit is capacitively coupled to a quarter-wave readout resonator that couples inductively to a bandpass (Purcell) filtered feedline. (d) The resonator frequencies \(\omega_{\rm Res}/2\pi\) are near \SI{7}{\giga\hertz} with \(\chi/\kappa_{\rm eff}\) ratios ranging from \(0.12\) to \(0.19\), where \(\chi\) and \(\kappa_{\rm eff}\) are respectively the dispersive shift and the effective resonator decay rate through the feedline. 
	Table of the qubit lifetimes (\(T_1\)) and operating frequencies (\(\omega_{\rm Qubit}/2\pi\)). Qubit color indicate the qubit operating frequency: red (purple) \(\rightarrow\) lowest (highest) operating frequency.}
\end{figure*}

In this work, we present machine-learning-enabled qubit-state discrimination. We evaluate the qubit-state discrimination performance of deep neural networks (DNN) relative to contemporary methods used for superconducting qubits. Nonlinear filters such as DNNs can better cope with system-dependent nonidealities, such as readout crosstalk. To evaluate these different qubit-state discriminator techniques, we use a quantum system comprising five frequency-tunable transmon qubits read out simultaneously via a common feedline using a standard frequency multiplexing approach. In contrast to single-qubit readout, such a multi-qubit system is subject to nonidealities, such as readout crosstalk, that may benefit from more sophisticated discriminators. We show that a DNN classifier can efficiently converge to a higher-performing multi-qubit discriminator with sufficient training. In our five-qubit system, we show that qubit-state assignment errors are reduced by up to \SI{25}{\percent} for multi-qubit architectures sharing a readout transmission line~\cite{Heinsoo2018_fast_readout, Arute2019_supremacy, Bultink2020_leak}. By examining the qubit-state assignment performance using a confusion matrix and the cross-fidelity metric, we attribute the reduction to the DNN compensating for crosstalk.

\ReMo{For systems with multiple superconducting qubits, readout crosstalk is a combination of (1) interactions between the generated readout probe signals, (2) photon population due to a residual coupling to a probe tone or neighboring readout resonators, (3) coupling between readout resonator and neighboring qubits, and (4) interactions between reflected/transmitted readout signals in the amplifier chain, mixers, or during analog demodulation and digitization. Fast readout, such as necessary for ancilla qubits as part of a quantum error correction protocol, requires wide resonator linewidths \(\kappa\). The frequency spacing between readout resonators is constrained by the qubit transition frequency, the number of frequency-multiplexed probe tones, and the readout amplifier chain bandwidth. Readout crosstalk is proportional to the spectral overlap between resonators, and thus, the wider the resonator linewidths, the more readout crosstalk. Therefore, readout crosstalk is expected to be a particularly significant error source for fast frequency-multiplexed ancilla qubit readout.}

It has been shown that neural networks can learn the quantum evolution of a single superconducting qubit using merely measurement data and without introducing the rules of quantum mechanics~\cite{Flurin2020_RNN}. Statistical learning algorithms have been applied to superconducting qubit readout in the form of support vector machines~\cite{Magesan2015_SVM}, hidden Markov models~\cite{Martinez2020_HMM}, or a reservoir computing approach~\cite{angelatos2020reservoir}. Using DNNs, improved single-qubit readout fidelity has previously been demonstrated for trapped-ions and spin qubits~\cite{Seif2018_NN, Ding2019_NN, matsumoto2020_DNN}. In this manuscript, we extend the application of neural networks to superconducting qubit readout and, more generally, to dispersive qubit readout. Furthermore, we demonstrate readout discrimination using a DNN of multiple simultaneously read out qubits on a single feedline. While we apply our methods to a superconducting qubit system, we anticipate that they will generalize to other platforms. 

\section{\label{sec:SQR}Superconducting Qubit Readout}


\begin{figure*}[t]
	\centering
	\includegraphics[width=0.94\textwidth]{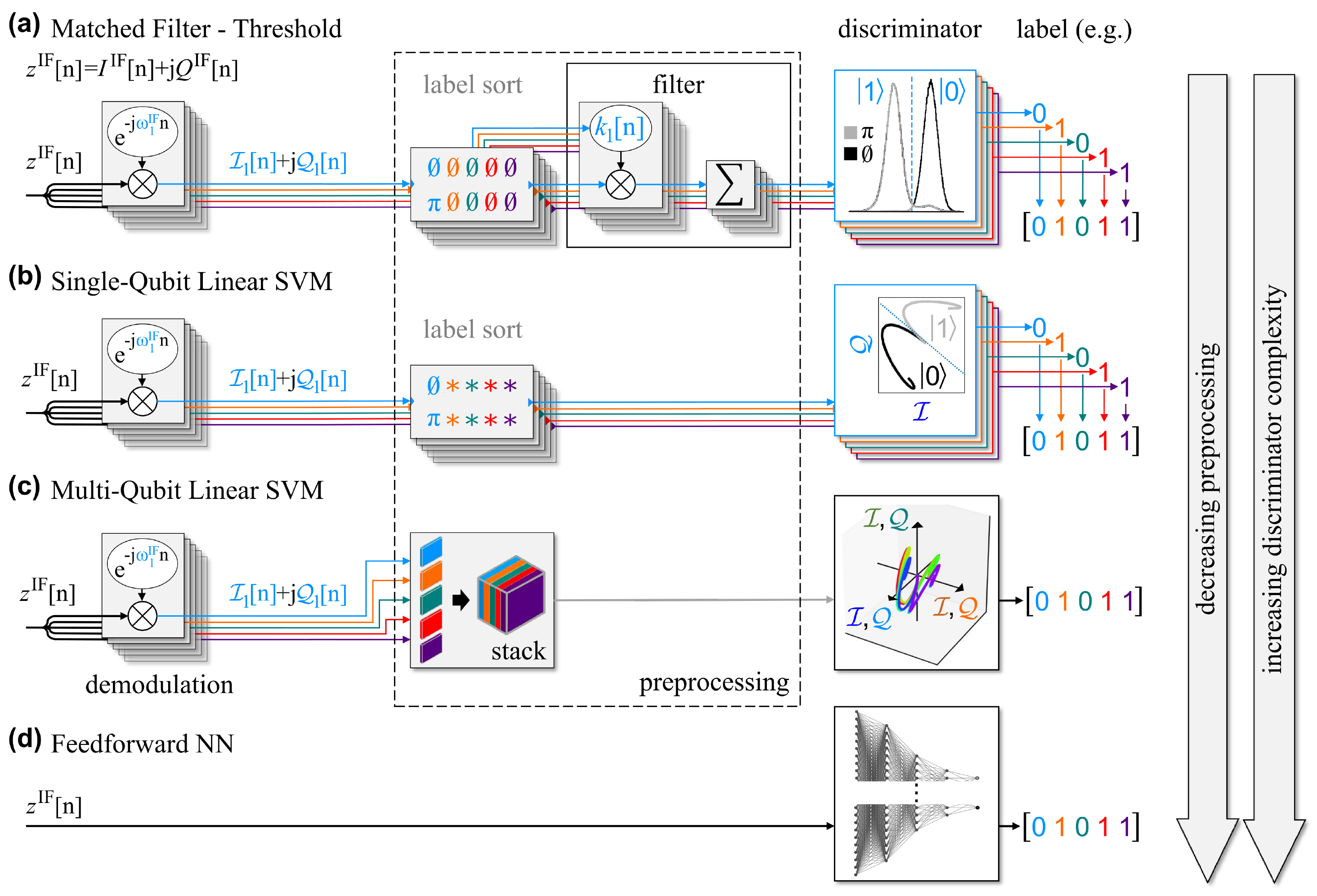}
    \caption{\label{fig:NNM}Measurement Data Processing and Discrimination. 
    (a) Superconducting qubit-state discrimination can be accomplished using a single-qubit matched filter (MF) with kernel \(k_i\text{[n]}\) which serves as a windowing function that projects the readout signals to a single axis and subsequent discriminator threshold optimization (no pulse applied, denoted by \(\emptyset\), qubit initialized in the ground state: \(\emptyset\rightarrow\ket{0}\) and labeled as \(0\); \(\pi\)-pulse applied, denoted by \(\pi\), qubit initialized in the excited state: \(\pi\rightarrow\ket{1}\) and labeled as \(1\)). We analyze (b) single-qubit linear support vector machines (SQ-LSVM), (c) multi-qubit LSVMs (MQ-LSVM), and (d) fully-connected feedforward neural networks (NN) as alternatives to MFs. The qubit-state-assignment fidelity of the MF and LSVM is maximized if the intermediate frequency signal (\(z^{\rm IF}\text{[n]}=I^{\rm IF}\text{[n]}+\text{j}Q^{\rm IF}\text{[n]}\)) is digitally demodulated (e.g., for resonator 1: \(z^{\rm IF}\text{[n]}{.\ast}^{\rm -j\omega_1^{\rm IF}n}=\mathcal{I}_1\text{[n]}+\text{j}\mathcal{Q}_1\text{[n]}\) with \({.\ast}\) indicating an element-wise multiplication). The training data is relabelled to train five parallel single-qubit discriminators (MF, SQ-LSVM). The training data can either be limited to measurements during which spectator qubits are kept in their ground state (denoted by \(\emptyset\)) or in all combinations of the ground and excited state (symbolized by \(\ast\). 
    The MQ-LSVM as a single multi-qubit discriminator requires the digitally demodulated data to be stacked and concatenated to form a single data block. The feedforward NN does not require any digital demodulation or preprocessing.
    }
\end{figure*}

Superconducting qubit readout is generally performed today under the paradigm of circuit quantum electrodynamics (cQED) in the dispersive regime~\cite{Blais2004}. Here, the qubit is coupled to a far-detuned resonator, such that their interaction can be treated perturbatively. The leading-order effect on the resonator is a qubit-state-dependent frequency shift \(\hat{H}_{\rm disp} = \chi\hat{a}^{\dagger}\hat{a}\hat{\sigma}_z\),
where $\hat{a}$ is the resonator lower operator, $\hat{\sigma}_z$ the Pauli-Z operator describing the qubit state, and $\chi$ the dispersive frequency shift. As a result, a coherent microwave signal incident on the resonator acquires a qubit-state-dependent phase shift upon transmission or reflection. The readout resonator population has to remain below a critical photon number, typically tens to hundreds of photons, to remain in the dispersive readout regime. Low-noise cryogenic preamplification---a Josephson traveling-wave parametric amplifier (JTWPA)~\cite{Macklin2015_TWPA} at the mixing chamber (\SI{20}{\milli\kelvin}) and a high-electron-mobility transistor (HEMT) at \SI{3}{\kelvin}--- are used to improve the signal-to-noise ratio (SNR). 
Subsequent heterodyne detection and digitization of the amplified signal imprints the information of the qubit state in the in-phase (\textit{I}) and quadrature (\textit{Q}) components of the output signal, as depicted in Fig.~\ref{fig:SCQC}(a). 

For multi-qubit systems, there are three main qubit-state-readout approaches. First, each qubit can be measured with a separate readout resonator, feedline, and amplifier chain---a resource-intensive approach with minimal crosstalk. Alternatively, more-resource-efficient readout architectures have several qubits coupled to a single readout resonator~\cite{DiCarlo2010_CavityShare} or use frequency-multiplexed readout signals from multiple readout resonators~\cite{Jerger2012_Multiplex} sharing a single feedline and amplifier chain~\cite{Evan2014_StateMeas}. In many contemporary architectures, Purcell filters are added to further reduce residual off-resonant energy decay from the qubits to the resonators~\cite{Sete2015_Purcell,Neill2018_Blueprint}.

For a qubit with static coupling to its readout resonator, energy decay and excitation during the readout are typically the primary sources of qubit measurement errors. 
In addition, a frequency-multiplexed readout signal contains state information on multiple qubits and is susceptible to crosstalk-induced qubit-state-readout errors. 
Such crosstalk errors occur due to intrinsic interactions between the qubits themselves, qubits coupling parasitically to the readout resonators associated with other qubits, or insufficient spectral separation between readout frequencies~\cite{Heinsoo2018_fast_readout}.

As a result of crosstalk, state transitions due to decoherence, and other nonidealities~\cite{Govia2015}, multi-qubit heterodyne signals are more complicated than for single qubits, making state discrimination more challenging. 
There has been significant progress in reducing error rates and measurement times for both single- and multi-qubit devices~\cite{Walter2017_fast_readout, Heinsoo2018_fast_readout}. However, managing, classifying, and extracting useful information from the measured signal remains an important challenge in light of the complex error mechanisms, such as crosstalk, introduced by multiplexed readout at scale. 

Here, we focus on multiple frequency-tunable transmon qubits~\cite{Koch2007_Transmon} arranged in a linear array with operating frequencies \(\omega_{\rm Qubit}/2\pi\) between \SI{4.3}{\giga\hertz} and \SI{5.2}{\giga\hertz} and qubit lifetimes \(T_1\) ranging from \SIrange{7}{40}{\micro\second} (see Appendix~\ref{sec:CHI} for additional details). The qubits are connected via individual co-planar waveguide resonators to the same Purcell filtered feedline, as depicted in Fig.~\ref{fig:SCQC}(b,c). The frequency-multiplexed readout tone comprises superposed baseband signals at intermediate frequencies (IF) between \SIrange{10}{150}{\mega\hertz} up-converted to the individual readout resonator frequencies \(\omega_{\rm Res}\). After passing the feedline, the transmitted and phase-shifted tones are down-converted to IF. 
Up- and down-conversion is conducted with a shared local oscillator at \SI{7.127}{\giga\hertz}. Lastly, the down-converted \textit{I}- and \textit{Q}-components of the signal are digitized with a \SI{2}{\nano\second} sampling period. The resulting sequences, \(I^{\rm IF}[n]\) and \(Q^{\rm IF}[n]\), are subsequently digitally processed---the focus of this work---to extract the individual qubit states. 

\section{\label{sec:QSD}Qubit-State Discrimination}

We employ supervised machine learning methods to improve superconducting qubit-state readout. This requires a classifier capable of distinguishing the qubit-state-dependent phase shift encoded in the discrete-time \(I^{\rm IF}\)[n] and \(Q^{\rm IF}\)[n] sequences. This section will also review the current approaches to state discrimination (which we will use as comparative benchmarks).


\textbf{Boxcar filters} average the equal-weighted digitally-demodulated elements of the \(I^{\rm IF}\)[n] and \(Q^{\rm IF}\)[n] discrete-time readout signal. The digital demodulation employed here is further elaborated in Appendix~\ref{sec:REA}. Each boxcar filtered digitally-demodulated sequence \(\mathcal{I}\)[n] and \(\mathcal{Q}\)[n] results in a single two-dimensional data point in the \(\mathcal{IQ}\)-plane~\cite{Krantz2019_review}. Subsequently, the resulting data set can be further processed and discriminated such as for example with a support vector machine (see Appendix~\ref{sec:REA}).


\textbf{Matched filter (MF)} windows are generalized windowing functions with each element optimized to maximize the SNR within a given system noise model~\cite{Turin1960_MF}. The boxcar window is the simplest example of a filter in the absence of such a noise model. For additive stationary noise independent of the qubit state and diagonal Gaussian covariance matrices, the optimal filter in terms of the SNR uses a ``window'' or ``kernel,'' proportional to the difference between the mean ground- and excited-state-readout signal, referred to as a ``matched filter'' in Ref.~\cite{Ryan2015_MF}, ``mode matched filter'' in Ref.~\cite{Heinsoo2018_fast_readout}, or as ``Fisher's linear discriminant'' in the context of statistics and machine learning~\cite{Bishop2006_Patt}. Applying such a matched filter reduces each readout single-shot measurement to a single one-dimensional value dependent on the qubit-state-dependent phase, allowing the qubit states to be discriminated by a simple threshold classifier. Here, we refer to a discriminator composed of a matched filter~\cite{Ryan2015_MF} and subsequently optimized threshold as MF.

While MFs are computationally efficient and provably optimal (for stationary noise) for single qubits, the computational complexity to derive multi-qubit MFs scales exponentially in the number of qubits, N~\cite{Fukunaga1990_MF}. Consequently, in practice, multi-qubit readout is conducted per qubit with individually optimized single-qubit MFs---the approach used for many contemporary single- and multi-qubit readout schemes~\cite{Ryan2015_MF, Heinsoo2018_fast_readout, Bronn2017_MF, Bultink2018_QE, Arute2019_supremacy} and does not account for noise sources and nonidealities present in mulit-qubit systems.

The MF kernel \(k_i\text{[n]}\) is equal to the difference between the mean ground- and excited-state readout signal normalized by its standard deviation, which must be measured experimentally using calibration runs with known qubit states. In our setup, the highest qubit-state-assignment fidelity for MFs is achieved using time traces recorded with the other qubits (spectator qubits) initialized in their ground states, as depicted in Fig.~\ref{fig:NNM}(a). 
This is a consequence of the simple noise model presumed for the MF, and thus, the MF discriminator does not capture multi-qubit readout crosstalk. In this paper we use the MF as a baseline to compare the following methods (see the Appendix~\ref{sec:REA} for  other variations of all the methods).

\textbf{Support vector machines (SVM)} are quadratic programs~\cite{Boser1992_SVM, Cortes1995_SVM} with the objective to maximize the distance between each data point and a decision boundary, a learned hyperplane separating two distinct classes. SVMs are a purely geometric approach to discrimination. For a single superconducting qubit, it has been reported that SVMs generate decision boundaries superior to that of MFs, as realistic noise deviates from the simple single-qubit noise model assumed for the MF~\cite{Magesan2015_SVM}.

Similar to the MF approach, multi-qubit-state discrimination can be conducted using a SVM classifier per qubit-readout signal. 
In contrast to our MF tune-up, we find that the highest assignment fidelity is achieved when the SVMs are trained using qubit-state measurement traces with the spectator qubits prepared in all combinations of ground and excited states. 

Alternatively, multi-qubit states can be discriminated by a single SVM composed of several hyperplanes that partition the full multidimensional \(\mathcal{IQ}\)-space, shown in Fig.~\ref{fig:NNM}(c). Such a multi-qubit SVM can be tuned using a ``one-versus-all'' strategy. We solve \(2^{N}\) (\textit{N}, the number of qubits) two-class discrimination problems with a single qubit state as one class and the remaining qubit states as the other. In our analysis, linear SVMs (LSVM) used as parallel single- and multi-qubit discriminators outperform their nonlinear counterparts in robustness, computational efficiency, and assignment fidelity~(see Appendix~\ref{subsec:SVM}). 

\begin{figure*}[t]
	\centering
	\includegraphics[width=0.94\textwidth]{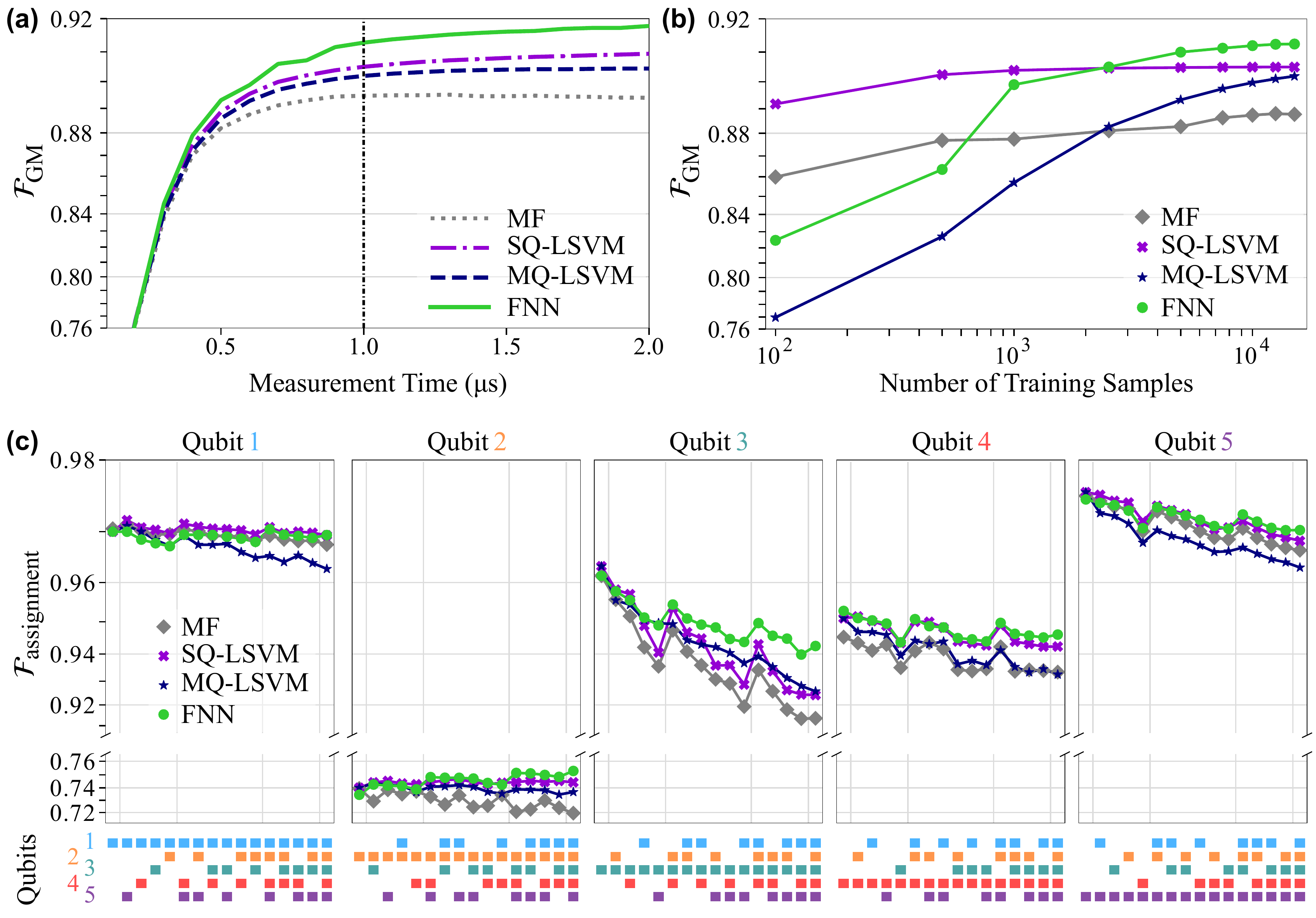}
    \caption{\label{fig:NNR}Qubit-State-Assignment Fidelity. (a) Geometric mean qubit-state-assignment fidelity \(\mathcal{F}_{\rm GM}\) (Eq.~\ref{eq:Fgeo}) 
    for five qubits versus measurement time for the matched filter (MF), single-qubit linear support vector machine (SQ-LSVM), multi-qubit linear SVM (MQ-LSVM), and the fully-connected feedforward neural network (FNN). (b) \(\mathcal{F}_{\rm GM}\) versus the number of training instances for each of the 32 qubit-state configurations evaluated after a measurement time of \SI{1}{\micro\second} [vertical dashed-dotted line in (a)]. (c) Achievable assignment fidelity \(\mathcal{F}_{\rm assignment}\) per qubit when \(N=\{1,2,\dots,5\}\) qubits are simultaneously discriminated after a \SI{1}{\micro\second}-measurement time. For each \(N\)-qubit discrimination task, the spectator qubits are initialized in their ground state. Single-qubit discrimination (\(N=1\)): the first data point of each of the five panels represents the single-qubit \(\mathcal{F}_{\rm assignment}\) defined by Eq.~\ref{eq:Fid}, while the states of the four spectator qubits are not discriminated and initialized in their ground state. When employed as single-qubit discriminators, all methods perform similarly. Two-qubit discrimination (\(N=2\)): The following four data points show \(\mathcal{F}_{\rm assignment}\) when the state of each panel's qubit is simultaneously discriminated with the state of one other qubit. N-qubit discrimination (\(N>2\)): the state of each panel's qubit is simultaneously discriminated with the states of \(N-1\) other qubits. For each \(N\)-qubit discrimination task, the non-spectator qubits are indicated with a colored square at the graph bottom.}
\end{figure*}

\textbf{Deep neural networks (DNN)} are mapping functions composed of arbitrarily connected nodes arranged in layers~\cite{Goodfellow2016_DL}. Depending on the layer organization and the functions governing the connections between nodes, different neural network archetypes can be generated. Here, we investigate three of the most common and successful DNNs: fully-connected feedforward neural networks, convolutional neural networks, and recurrent neural networks. We find a fully-connected feedforward neural network (FNN)---implemented in PyTorch~\cite{Paszke2019_pytorch}---outperforms the other network architectures in qubit-state-assignment fidelity. Our FNN architecture is composed of three hidden layers (1st, 2nd, and 3rd layer consist of 1000, 500, and 250 nodes, respectively) that use SELU activation functions~\cite{Klambauer2017_SELU}, and a softmax 
applied to the \(2^{N}\)-node output layer. The network is trained (validation-training set ratio of 0.35) using the Adam optimizer~\cite{kingma2017_adam} with categorical cross-entropy as the loss function.

In contrast to the MF and LSVM, the FNN can directly discriminate the frequency-multiplexed multi-qubit readout sequences \(I^{\rm IF}\)[n] and \(Q^{\rm IF}\)[n] without demodulation or filtering \ReMo{(see Appendix~\ref{sec:REA} for additional information and results)}. Training the network directly on the multiplexed readout signal bypasses the need for further preprocessing stages, suggesting a more efficient use of the measurement output, as illustrated in Fig.~\ref{fig:NNM}(d). In addition, fewer independent operations in the readout chain may reduce the possibility of systematic errors. 

\section{\label{sec:RES}Results}

We now present our five-qubit readout experiment results, comparing the performance of parallelized single-qubit MFs, parallelized single-qubit LSVMs (SQ-LSVM), multi-qubit LSVM (MQ-LSVM), and FNN approaches. The same qubit-readout sequences \(I^{\rm IF}\)[n] and \(Q^{\rm IF}\)[n] with varying amounts of preprocessing [Fig.~\ref{fig:NNM}]---are used for all approaches. We compare the discrimination results, a five-bit string with each bit representing the assigned state of a qubit. The qubit-state-assignment fidelity for qubit \(i\) is 
\begin{equation}
    \mathcal{F}_{i}=1-[P(0_i\vert\pi_i)+P(1_i\vert\emptyset_i)]/2,
    \label{eq:Fid}
\end{equation}
where \(P(0_i\vert\pi_i)\) is the conditional probability of assigning the ground state with label \(0\) to qubit \(i\) when prepared in the excited state with a \(\pi\)-pulse applied. \(P(1_i\vert\emptyset_i)\) is the conditional probability of assigning the excited state with label \(1\) to qubit \(i\) when prepared in the ground state (no pulse applied: \(\emptyset\)).

The data to train and evaluate the discriminator performance was acquired using the five-qubit chip introduced in Fig.~\ref{fig:SCQC}(b,c). For five qubits, all 32 qubit-state permutations are sequentially initialized and the measurement output is recorded. The generated data set contains 50,000 single-shot sequences \(I^{\rm IF}\)[n] and \(Q^{\rm IF}\)[n] recorded over \SI{2}{\micro\second} for each qubit-state configuration. The recorded data set is subsequently divided into a randomized training and test set (15,000 traces per qubit-state configuration for training and 35,000 for testing). All of the following results are evaluated using 35,000 single-shot measurements per qubit-state configuration.

\begin{table*}
\caption{\label{tab:QUF}Qubit-assignment fidelity if discriminated individually, \(\mathcal{F}^{\rm 1Q}_i\), and in parallel with all other qubits, \(\mathcal{F}^{\rm 5Q}_i\). The last five columns present the assignment fidelity for an \(N\)-qubit discrimination process with \(N=\{1,2,\dots,5\}\). 
\(\langle\mathcal{F}^{\rm \textit{N}Q}\rangle\) represents the mean assignment fidelity of all qubit permutations. The single-qubit assignment fidelity is similar for all discriminator approaches. For a two-qubit discrimination task, the SQ-LSVM and FNN outperform the MF and MQ-LSVM. For \(N\)-discrimination tasks with \(N>2\), the FNN outperforms all other methods.}
\begin{tabular}{@{}l|cccccccccc|ccccc@{}}
\midrule\midrule
&\multicolumn{2}{c}{Qubit \color{Q1}{1}}&\multicolumn{2}{c}{Qubit \color{Q2}{2}}&\multicolumn{2}{c}{Qubit \color{Q3}{3}}&\multicolumn{2}{c}{Qubit \color{Q4}{4}}&\multicolumn{2}{c|}{Qubit \color{Q5}{5}}
&\multirow{2}{*}{~~\(\langle\mathcal{F}^{\rm 1Q}\rangle\)}&\multirow{2}{*}{\(\langle\mathcal{F}^{\rm 2Q}\rangle\)}&\multirow{2}{*}{\(\langle\mathcal{F}^{\rm 3Q}\rangle\)}&\multirow{2}{*}{\(\langle\mathcal{F}^{\rm 4Q}\rangle\)}&\multirow{2}{*}{\(\langle\mathcal{F}^{\rm 5Q}\rangle\)~~}\\
&\(\mathcal{F}^{\rm 1Q}_{\color{Q1}{1}}\)&\(\mathcal{F}^{\rm 5Q}_{\color{Q1}{1}}\)&\(\mathcal{F}^{\rm 1Q}_{\color{Q2}{2}}\)&\(\mathcal{F}^{\rm 5Q}_{\color{Q2}{2}}\)&\(\mathcal{F}^{\rm 1Q}_{\color{Q3}{3}}\)&\(\mathcal{F}^{\rm 5Q}_{\color{Q3}{3}}\)
&\(\mathcal{F}^{\rm 1Q}_{\color{Q4}{4}}\)&\(\mathcal{F}^{\rm 5Q}_{\color{Q4}{4}}\)&\(\mathcal{F}^{\rm 1Q}_{\color{Q5}{5}}\)&\(\mathcal{F}^{\rm 5Q}_{\color{Q5}{5}}\)&\\\midrule
MF      &~0.971  & 0.968~    &~0.740 & 0.719~     &~0.962 & 0.914~    &~0.946 & 0.934~    &~0.976 & 0.967~    &~0.9185  & 0.9100  & 0.9042  & 0.8993  & 0.8946 \\
SQ-LSVM &~0.970  & 0.969~    &~0.740 & 0.744~     &~0.963 & 0.924~    &~0.951 & 0.943~    &~0.976 & 0.968~    &\textbf{0.9201}  & \textbf{0.9148}  & 0.9112  & 0.9083  & 0.9053\\
MQ-LSVM &~0.970  & 0.963~    &~0.740 & 0.737~     &~0.963 & 0.926~    &~0.951 & 0.933~    &~0.976 & 0.963~    &~0.9201  & 0.9130  & 0.9078  & 0.9033  & 0.8997\\
FNN     &~0.970  & \textbf{0.969}    &~0.735 & \textbf{0.753}~     &~0.962 & \textbf{0.943}~    &~0.953 & \textbf{0.946}~    &~0.975 & \textbf{0.970}    &~0.9188  & 0.9141  & \textbf{0.9129}  & \textbf{0.9126}  & \textbf{0.9122} \\
\midrule\midrule
\end{tabular}
\end{table*}

We quantify the assignment fidelity per qubit using the geometric mean assignment fidelity, 
\begin{equation}
    \mathcal{F}_{\rm GM}=(\mathcal{F}_{\color{Q1}{1}}\mathcal{F}_{\color{Q2}{2}}\mathcal{F}_{\color{Q3}{3}}\mathcal{F}_{\color{Q4}{4}}\mathcal{F}_{\color{Q5}{5}})^{1/5}, 
    \label{eq:Fgeo}
\end{equation}
\noindent
with each qubit-state-assignment fidelity defined by Eq.~\ref{eq:Fid}.
Both SVM approaches improve the assignment fidelity relative to the MF, with the parallelized single-qubit SVM outperforming the multi-qubit approach by \SI{0.3}{\percent} after a \SI{1}{\micro\second}-measurement time. For multi-class discriminators such as the MQ-LSVM, geometric constraints result in ambiguous regions without a unique class assigned~\cite{DudaHart1973_MQSVM}, which leads to poor performance relative to the other approaches. After a \SI{1}{\micro\second}-long measurement time, the FNN, compared to the MF, increases the qubit-state-assignment fidelity from \(0.885\) to \(0.913\)---a reduction of the single-qubit assignment error [\(1-(1-\mathcal{F}_{\rm FNN})/(1-\mathcal{F}_{\rm MF})\)] by 0.244. Compared to the SQ-LSVM, the FNN increases the qubit-state-assignment fidelity from \(0.905\) to \(0.913\) and thus reduces the single-qubit assignment error by \(0.084\). The FNN yields the highest qubit-state-assignment fidelity regardless of measurement time [Fig.~\ref{fig:NNR}(a)]. See Appendix~\ref{sec:REA} for additional comparison of discriminators and data processing methods.

\begin{figure}[htbp!]
	\centering
	\includegraphics[width=0.475\textwidth]{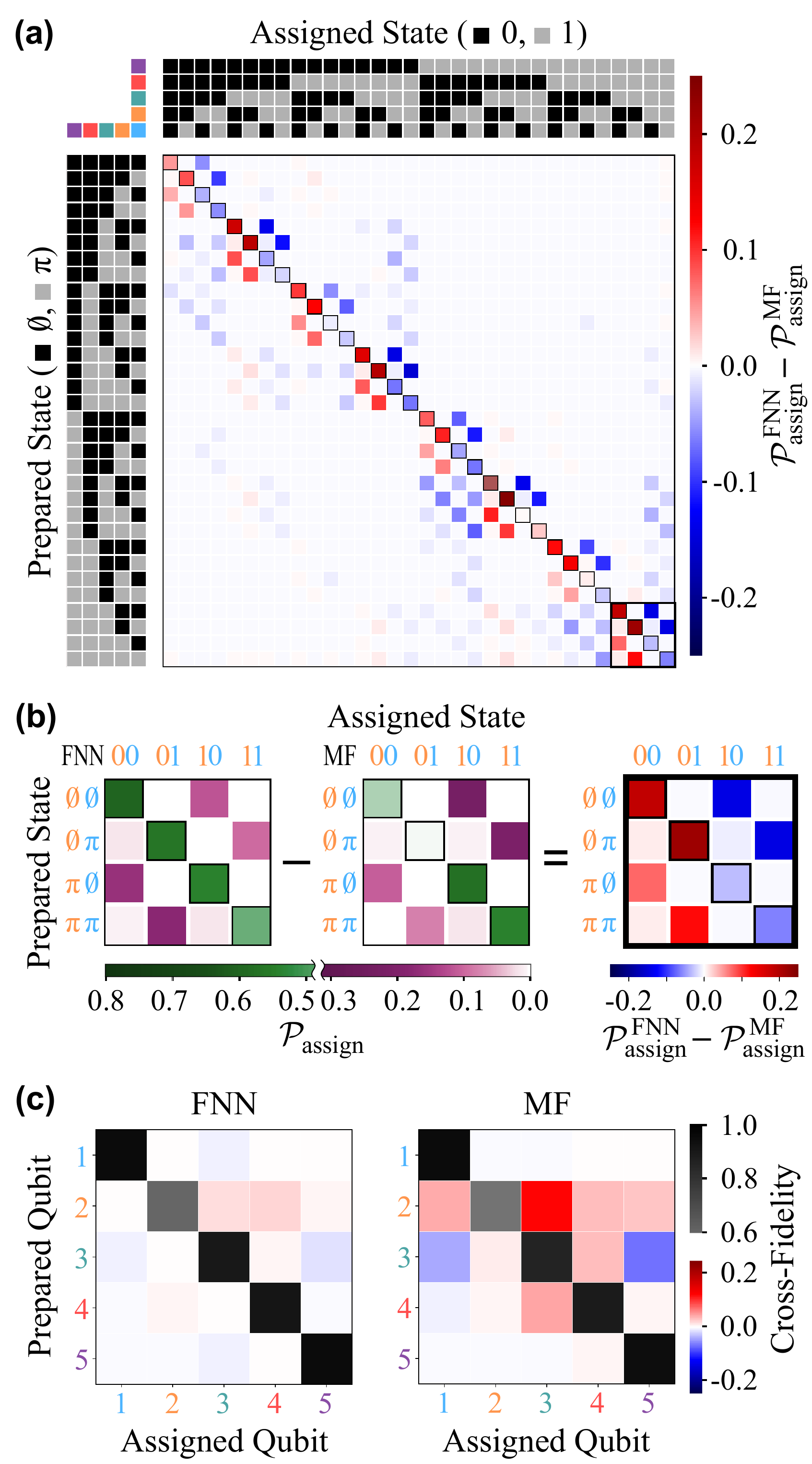}
	\caption{\label{fig:NNC}Assignment Fidelity Analysis. (a) Difference between the confusion (assignment probability) matrix of the feedforward neural network (FNN) \(\mathcal{P}_{\rm assign}^{\rm FNN}\) and of the matched filter (MF) \(\mathcal{P}_{\rm assign}^{\rm MF}\). The rows of the confusion matrix encompass the discriminator's probability distribution to assign each of the 32 qubit-state configurations to the row's prepared qubit-state configuration (no pulse applied, qubit initialized in the ground state: \(\emptyset\rightarrow 0\); \(\pi\)-pulse applied, qubit initialized in the excited state: \(\pi\rightarrow 1\)). An increase (decrease) in the relative state-assignment probability is marked in red (blue). Red diagonal and blue off-diagonal elements indicate an improvement of the FNN over MF discrimination performance. (b) The cutouts [bold frame in the lower right corner of (a)] of the FNN, MF, and resulting relative confusion matrix display the most prominent pattern that arises due to the discrimination of qubit 2. (c) The FNN and MF cross-fidelity matrices, as defined in Eq.~\ref{eq:CFid}, indicate the discrimination correlation. 
	The off-diagonals are ideally \(0\). A positive (negative) matrix off-diagonal entry indicates qubit-state assignment to be correlated (anti-correlated).}
\end{figure}

Next, we evaluate the assignment fidelity for different numbers of training samples per qubit configuration, presented in Fig.~\ref{fig:NNR}(b). The assignment fidelity of five parallel single-qubit discriminators (MF, SQ-LSVM) saturates around 1,000 training samples per qubit-state configuration. The assignment fidelity of the FNN exceeds that of parallelized single-qubit discriminators after 2,500 training samples and saturates around 10,000 training samples per qubit-state configuration. We estimate that the multi-qubit LSVM plateaus after approximately 40,000 training samples per qubit-state configuration. The FNN architecture here is solely optimized to maximize the qubit-state-assignment fidelity, with no consideration of the size of training data required. Thus, these results should not be taken as an indication that DNN approaches will generically perform poorly for small training sets. The remaining discriminator analysis is conducted after a \SI{1}{\micro\second}-measurement time and 10,000 training samples per qubit-state configuration. 

The assignment fidelity per qubit, discriminated individually and in parallel with up to \(N=5\) qubits, is presented in Fig.~\ref{fig:NNR}(c). For \(N\)-qubit discrimination tasks with \(N>2\), the FNN starts outperforming its discriminator alternatives. Except for qubit 2, the per-qubit-assignment fidelity decreases with an increasing number of discriminated qubits. We observe a more substantial assignment fidelity decrease if the resonators involved in the discrimination are proximal in frequency, suggesting the occurrence of readout crosstalk. In addition to readout crosstalk, qubit 3 reveals control crosstalk with qubit 1 and 5, the qubits closest in frequency. Under the assumption of additive stationary noise independent of the qubit state and diagonal Gaussian covariance matrices, the estimated upper qubit-state-assignment fidelity bound per qubit for MFs~\cite{Magesan2015_SVM} including the label confidence are \(\mathcal{F}^{\rm \overline{MF}}_{\color{Q1}{1}}\approx 0.974\), \(\mathcal{F}^{\rm \overline{MF}}_{\color{Q2}{2}}\approx 0.773\), \(\mathcal{F}^{\rm \overline{MF}}_{\color{Q3}{3}}\approx 0.965\), \(\mathcal{F}^{\rm \overline{MF}}_{\color{Q4}{4}}\approx 0.95\), and \(\mathcal{F}^{\rm \overline{MF}}_{\color{Q5}{5}}\approx 0.979\), respectively (see Appendix~\ref{subsec:MF} for additional details). \(\mathcal{F}^{\rm \overline{MF}}_{\color{Q2}{2}}\) is primarily reduced due to \(T_1\)-events and limited qubit-state separation in the \(\mathcal{IQ}\)-plane. The different discriminators yield a similar assignment fidelity within a few tenths of a percent of the upper MF assignment fidelity bound---except for qubit 2 where it is off by a few percent---when tasked to discriminate a single qubit, as shown in Tab.~\ref{tab:QUF}. The small discrepancy between this upper bound and the achieved assignment fidelity suggests that the noise sources affecting single-qubit readout in our devices are reasonably well approximated by additive stationary noise independent of the qubit state and diagonal Gaussian covariance matrices. 
As the number of simultaneously discriminated qubits increases, the assignment fidelity increasingly deviates from \(\mathcal{F}^{\rm \overline{MF}}_{i}\), revealing system dynamics unaccounted for by the Gaussian noise model. 

The confusion matrix, a matrix \(\mathcal{P}_{\rm assign}\) with the qubit-state-assignment probability distribution for each prepared qubit-state configuration as rows, provides further insight into the underlying error mechanisms. The confusion matrix is the identity matrix if each prepared state is correctly labeled and assigned. In practice, in addition to misclassification, the preparation of states can be imperfect. 
We estimate the mean state preparation fidelities for each qubit (see Appendix~\ref{subsec:MF}): \(\mathcal{F}^{\rm prep}_{\color{Q1}{1}}\approx 0.995\), \(\mathcal{F}^{\rm prep}_{\color{Q2}{2}}\approx 0.986\), \(\mathcal{F}^{\rm prep}_{\color{Q3}{3}}\approx 0.977\), \(\mathcal{F}^{\rm prep}_{\color{Q4}{4}}\approx 0.976\), and \(\mathcal{F}^{\rm prep}_{\color{Q5}{5}}\approx 0.985\). 

The qubit-state-dependent assignment probability of our FNN relative to the MF is expressed as the difference between their respective confusion matrices, \(\mathcal{P}_{\rm assign}^{\rm FNN}\) and \(\mathcal{P}_{\rm assign}^{\rm MF}\), shown in Fig.~\ref{fig:NNC}(a). The FNN generally reduces the erroneous off-diagonal assignment probabilities relative to the MF. The most significant exception being the lower off-diagonal elements corresponding to decay of qubit 2, as presented in Fig.~\ref{fig:NNC}(b).

Deviations from the ideal confusion matrix occur due to initialization errors, state transitions during the measurement, or readout crosstalk. Typically, the qubit-state misclassifications in the lower off-diagonal block outweigh those of the upper off-diagonal due to the greater likelihood of decay events at cryogenic temperatures. Here, for a \SI{1}{\micro\second}-long measurement, qubit 2---the qubit with the shortest lifetime---has a \SI{15}{\percent} probability of \(T_1\)-decay, such that for a significant portion of the training measurements with qubit 2 excited, the final state of qubit 2 is the ground state. 

As shown in Fig.~\ref{fig:NNC}(b), the FNN is more likely to assign a ground-state label to qubit 2 than an excited-state label, whereas the MF reveals the reverse trend. This suggests that the assignment probabilities of the FNN agree better with the expected error model. However, we can attribute the pattern of the MF assignment probability to a training bias. Since measurements with qubit 2 prepared in the excited state and corrupted by a \(T_1\)-decay have integrated signals similar to measurements with qubit 2 prepared in the ground state, the threshold optimizer overcompensates to correctly classify $T_1$-decay corrupted excited-state measurements at the cost of misclassification of ground-state measurements. This results in the misclassification pattern seen in Fig.~\ref{fig:NNC}(b) for \(\mathcal{P}_{\rm assign}^{\rm MF}\).


From the confusion matrix, we can further extract the probability distribution of the non-zero Hamming distance. This is the probability distribution describing the number of misassigned qubits per qubit-state configuration. The assignment errors of the FNN (MF) \ReMo{occur in} \(\SI{85.8}{\percent}\) (\(\SI{83.8}{\percent}\)) \ReMo{of the cases as} single-qubit, \(\SI{13.2}{\percent}\) (\(\SI{15.0}{\percent}\)) \ReMo{as} two-qubit, and \(\SI{0.8}{\percent}\) (\(\SI{1.0}{\percent}\)) \ReMo{as} three-qubit errors. The reduction of assignment errors for the FNN compared to the MF is not specific to a unique Hamming distance error, indicating a consistent reduction of crosstalk.


\begin{table}[t]
\caption{\label{tab:CFI}Mean absolute value, \(\langle\vert\cdot\vert\rangle\), of the qubit-state-assignment correlations between readout resonators \(i\) and \(j\) (\(i \neq j\)) extracted from the cross-fidelity matrix \(\mathcal{F}^{\rm CF}\) when using a MF or FNN discriminator.}
\begin{tabular}{@{}l|cccc@{}}
\midrule\midrule
&~~\(\langle \vert \mathcal{F}^{\rm CF}_{j = i\pm 1}\vert \rangle\)~~&~~\(\langle \vert \mathcal{F}^{\rm CF}_{j = i\pm 2}\vert \rangle\)~~&~~\(\langle \vert \mathcal{F}^{\rm CF}_{j = i\pm 3}\vert \rangle\)~~&~~\(\langle \vert \mathcal{F}^{\rm CF}_{j = i\pm 4}\vert \rangle\)~~\\\midrule
MF      & 0.020 & 0.015 & 0.006 & \(\sim\)0 \\
FNN~~   & 0.002 & 0.005 & 0.002 & \(\sim\)0 \\
\midrule\midrule
\end{tabular}
\end{table}

To further study crosstalk, we consider the cross-fidelity matrix, which describes correlations between the assignment fidelities of individual qubits~\cite{Heinsoo2018_fast_readout}. The cross-fidelity \(\mathcal{F}^{\rm CF}_{ij}\) is defined as
\begin{equation}
    \mathcal{F}^{\rm CF}_{ij}=\langle 1-\left[P\left(1_i\vert \emptyset_j\right)+P\left(0_i\vert \pi_j\right)\right]\rangle,
    \label{eq:CFid}
\end{equation}
where \(\emptyset_j\) (\(\pi_j\)) represent the preparation of qubit \(j\) in the ground (excited) state and \(0_i\) (\(1_i\)) the subsequent assignment to the ground (excited) state (\(\langle f \rangle\) denotes the mean value of a function \(f\)). A positive (negative) off-diagonal indicates a correlation (anti-correlation) between the two qubits. Such correlations can occur due to readout crosstalk. The off-diagonal entries for the FNN are all less than one percent, and are drastically reduced relative to the MF. Relative to the MF, the mean cross-fidelity, \(\langle \vert \mathcal{F}^{\rm CF}_{ij} \vert \rangle\), for nearest neighbors \((j = i\pm 1)\) is reduced by one order of magnitude from \(\langle \vert \mathcal{F}^{\rm MF~CF}_{j = i\pm 1} \vert \rangle=0.02\) to \(\langle \vert \mathcal{F}^{\rm FNN~CF}_{j = i\pm 1} \vert \rangle=0.002\). For neighboring readout resonators, the spectral overlap is maximized, and thus readout crosstalk most likely to occur. In general, relative to the MF, the FNN reduces the mean cross-fidelity for all \(j\neq i\), as presented in Tab.~\ref{tab:CFI}. 
The FNN's reduction of assignment correlations by up to one order of magnitude corroborates the claim of the FNN's diminishing readout-crosstalk-related discrimination errors. 

\section{\label{sec:con}conclusion}
We have demonstrated an approach to multi-qubit readout using neural networks as multi-qubit state discriminators that is more crosstalk-resilient than other contemporary approaches. We find that a fully-connected FNN increases the readout assignment fidelity for a multi-qubit system compared to contemporary methods. We observe that the FNN compensates system-nonidealities such as readout crosstalk more effectively relative to alternatives such as matched filters (MFs) or support vector machines (SVMs). The assignment error rate is diminished by up to \SI{25}{\percent} and crosstalk-induced discrimination errors are suppressed by up to one order of magnitude. The relative assignment fidelity improvement of the FNN over its contemporary alternatives grows as the number of simultaneously read out and multiplexed qubits increases. 

While FNNs are initially more resource-intensive in training, its re-calibration can be significantly more efficient due to transfer learning~\cite{bengio2012_TL}. Periodic re-calibration of control and readout parameters is necessary as quantum systems drift in time. For a marginal drift, neural networks can be updated at a fraction of the initial resource requirements. 
Furthermore, to speed up qubit readout, the techniques developed here can be transitioned to dedicated hardware such as field-programmable gate arrays (FPGA)~\cite{Ding2019_NN}.

We have tested our FNN multi-qubit-state discrimination approach on a quantum system with five superconducting qubits and frequency-multiplexed readout. While the readout fidelity of Qubit 2 was relatively marginal, four qubits revealed multi-qubit readout fidelities comparable with contemporary multi-qubit systems, albeit with measurement times around \SI{1}{\micro\second} (see Appendix~\ref{sec:REA} for additional details), much longer than the state of the art of \SI{100}{\nano\second} for single-qubit systems~\cite{Walter2017_fast_readout}. We demonstrated an improvement using FNN for all qubits. The next step is to test the performance of FNNs on higher-fidelity multi-qubit systems with measurement times below \SI{100}{\nano\second} to assess if the advantage is retained on already high-performing devices. 
FNNs offer a readout-state discrimination approach tailored to the underlying system. They can be readily employed to more general discrimination tasks than we have considered here, such as multi-level readout in a qudit architecture~\cite{Kurpiers2018_QUT, Elder2020_MLR, Yurtalan2020_MLR, Wang2021_MLR}. This work presents a potential building block to scaling quantum processors while maintaining high-fidelity readout.

\section*{Acknowledgements}
We want to express our appreciation for Mirabella Pulido and Chihiro Watanabe for administrative assistance. This research was funded in part by the DARPA Polyplexus grant No. HR00112010001; by the U.S. Army Research Office (ARO) Multidisciplinary University Research Initiative (MURI) W911NF-18-1-0218; and by the Department of Defense via Lincoln Laboratory under Air Force Contract No. FA8721-05-C-0002. The views and conclusions contained herein are those of the authors and should not be interpreted as necessarily representing the official policies or endorsements, either expressed or implied, of DARPA or the US Government.

\appendix

\section{\label{sec:MEA}Measurement Setup}
Qubit control and readout pulses---envelopes with cosine shaped rising and falling edges encompassing a plateau---are programmed in Labber. They are created using three---two for control and one for readout---Keysight M3202A PXI arbitrary waveform generators (AWG) with a sampling rate of \SI{1}{\giga\sample\per\second}. The in-phase (\textit{I}) and quadrature (\textit{Q}) components of the signals at \si{\mega\hertz} frequencies are up-converted to the qubit transition frequency using an \textit{IQ}-mixer and a local oscillator (LO) (Rohde and Schwarz SGS100A) per AWG. The control and readout tones are combined and sent to the qubit chip in the dilution refrigerator via a single microwave line attenuated by \SI{60}{\decibel}.

The qubit chip is mounted in a microwave package following design principles as reported in Refs.~\cite{Lienhard2019_pack,huang2020_pack}. A coil---centered above the qubit chip---is mounted in the device package. A global flux bias \(\Phi\) is applied through that coil to the superconducting quantum interference devices (SQUID) of the qubits using a Yokogawa GS200.

The readout signal, upon acquisition of a qubit-state-dependent phase shift, is first amplified using a Josephson traveling-wave parametric amplifier (JTWPA) with near quantum-limited performance over a bandwidth of more than \SI{2}{\giga\hertz} and a \SI{1}{\deci\bel} compression point of approximately \SI{-100}{\deci\bel{m}}~\cite{Macklin2015_TWPA}. An Agilent E8267D signal generator provides the pump tone for the JTWPA. The microwave line carrying the pump tone is attenuated by \SI{50}{\deci\bel} and fed into the JTWPA via a set of directional couplers and isolators located in the mixing chamber of the refrigerator. The signal is further amplified by a high-electron-mobility transistor (HEMT) amplifier that is thermally anchored to the \SI{3}{\kelvin} stage.

At room temperature, the readout signal is amplified, IQ-mixed with the LO at \SI{7.127}{\giga\hertz}, and fed into a heterodyne detector. The \textit{I}- and \textit{Q}-components of the readout signal are digitized with a Keysight M3102A PXI Analog to Digital Converter (ADC) at a sampling rate of \SI{500}{\mega\sample\per\second}. The subsequent digital signal processing to distinguish qubit states is the focus of this manuscript. 

\section{\label{sec:CHI}Five-Qubit Chip}
The quantum system  five superconducting qubits is fabricated on a (001) silicon substrate (\(>\)\SI{3500}{\ohm\centi\meter}) by dry etching a molecular-beam epitaxy (MBE) grown aluminum film in an optical lithography process before being diced into \(5\times5~\text{mm}^2\) chips, as described in~\cite{Yan2016_NatCom_flux}. 

The superconducting chip consists of coplanar waveguides and five frequency-tunable transmon qubits~\cite{Koch2007_Transmon}. The target qubit transition frequencies alternate between \SI{4.3}{\giga\hertz} and \SI{5.2}{\giga\hertz}. The qubits are detuned (\(\rightarrow\) operating frequency) to limit qubit-qubit and control crosstalk. The capacitive nearest-neighbor (next-nearest-neighbor) qubit-qubit coupling rate, \(J_{nn}\) (\(J_{nnn}\)), is designed (using COMSOL Multiphysics\textregistered) to be \(J_{nn}/2\pi\approx \SI{14}{\mega\hertz}\) (\(J_{nnn}/2\pi<\SI{1}{\mega\hertz}\)) and at the qubit operating frequency \(<\SI{0.3}{\mega\hertz}\) (\(<\SI{0.01}{\mega\hertz}\))~\cite{blais2020_cQED}. Each qubit couples capacitively to a quarter-wave resonator that couples inductively to a shared bandpass (Purcell) filtered feedline. Neighboring readout resonator frequencies differ by \(\sim\)\SI{50}{\mega\hertz}. The qubit and resonator operation parameters are included in Tab.~\ref{tab:QUB} and Tab.~\ref{tab:RES}. 

\begin{table}[t]
\caption{\label{tab:QUB}Chip comprising five superconducting frequency-tunable transmon qubits with alternating transition frequencies. A normalized magnetic flux bias \(\Phi/\Phi_0\) (magnetic flux quantum \(\Phi_0\)) detunes the qubits from their idling to their operating frequency. The qubit anharmonicities \(\alpha\) are in the moderate transmon regime. The qubit lifetimes \(T_1\), Ramsey coherence times \(T_{\rm 2R}\), and spin-echo relaxation times \(T_{\rm 2E}\) are measured at the qubit operating frequency.}
\begin{tabular}{@{}c|cc|c|c|ccc@{}}
\midrule\midrule
Qubit~~&\multicolumn{2}{c|}{\(\omega_{\rm Qubit}/2\pi\)}&~~~Bias~~~&~~\(\alpha/2\pi\)~~&~~~\(T_1\)~~&~~\(T_{\rm 2R}\)~~&~~\(T_{\rm 2E}\)~~\\ 

            & Idle & Biased~                                 &~\multirow{2}{*}{\(\left(\frac{\Phi}{\Phi_0}\right)\)}~&~\multirow{2}{*}{\(\left(\si{\mega\hertz}\right)\)}~ & \multicolumn{3}{c}{\multirow{2}{*}{(\si{\micro\second})}}\\ 
            &  \multicolumn{2}{c|}{(\si{\giga\hertz})}      & &  & & & \\\midrule
1           &~5.249                 & 5.092~                & 0.124                 & -212                & 40.8                  & 1.3                   & 7.4 \\
2           &~4.708                 & 4.404~                & 0.160                 & -216                & 6.4                   & 0.6                   & 4.1 \\
3           &~5.202                 & 5.000~                & 0.166                 & -204                & 21.4                  & 1.0                   & 7.2 \\
4           &~4.560                 & 4.309~                & 0.154                 & -214                & 11.8                  & 0.8                   & 5.4 \\
5           &~5.196                 & 5.165~                & 0.085                 & -200                & 23.4                  & 7.6                   & 31.8 \\
\midrule\midrule
\end{tabular}
\end{table}

\begin{table}[t]
\caption{\label{tab:RES}Chip comprising five superconducting readout resonators at bare resonance frequencies \(\sim\)\SI{7}{\giga\hertz}. Signals are up-converted from \si{\mega\hertz} intermediate frequencies (IF) utilizing a common local oscillator at \(\omega_{\rm LO}/2\pi=\SI{7.127}{\giga\hertz}\). Each resonator couples to a designated qubit with strength \(g\), leading to a dispersive shift \(\chi\). The effective resonator decay rate through the Purcell filter is \(\kappa_{\rm eff}\). The qubit-resonator interaction remains in the dispersive regime for readout resonator photon populations below the critical photon number \(n_{\rm crit}\).}
\begin{tabular}{@{}c|cc|ccc|c@{}}
\midrule\midrule
Resonator~&\(\omega_{\rm Res}/2\pi\)&\(\omega_{\rm IF}/2\pi\)&~~\(g/2\pi\)~~&~~\(\chi/2\pi\)~~&\(\kappa_{\rm eff}/2\pi\)~~&~~\(n_{\rm crit}\)\\\ 
     & ~(\si{\giga\hertz})~  &~(\si{\mega\hertz})~ & \multicolumn{3}{c|}{\(\left(\si{\mega\hertz}\right)\)} & \\\midrule
1    & 7.06                 & -65                    & 116.3   & 0.83  & 4.29      & 33.8    \\
2    & 7.10                 & -26                    & 143.3   & 0.51  & 4.25      & 55.3    \\
3    & 7.15                 & 24                    & 125.7   & 0.77  & 4.41      & 34.9    \\
4    & 7.20                 & 70                    & 133.1   & 0.49  & 3.33      & 56.9    \\
5    & 7.25                 & 127                   & 125.4   & 0.80  & 6.90      & 33.0    \\
\midrule\midrule
\end{tabular}
\end{table}

\section{\label{sec:DIS}Qubit-State Discriminators}
The study of computational algorithms with the ability to improve through experience is typically referred to as machine learning~\cite{Bishop2006_Patt}. These algorithms strive to identify patterns in sample data, called training data, and create an approximate model of an underlying decision process without explicit instructions. While many machine learning ideas are several decades old, they only recently became widely applicable due to the development of sufficient computational resources and are applied today in image processing~\cite{LeCun1998_CNN}, natural language processing~\cite{devlin2019_bert}, or playing advanced games such as chess~\cite{Silver2018_RL}.

Machine learning can be broadly divided into three categories: unsupervised, supervised, and reinforcement learning. Here, we focus on supervised learning methods that learn an input-output mapping function using a trusted set of input-output pairs (training set). Typically, the input-output pairs for training are acquired by the ``supervisor,'' hence the terminology. The quality of the learned mapping function can be probed utilizing an additional set of trusted input-output pairs (test set). The comparison of performance of a supervised learning method on the training set compared to the test set is referred to as generalization. 

\begin{figure*}[t]
\captionof{table}{\label{tab:MF}Numerical values extracted from Gaussian fits to readout data distribution after a \SI{1}{\micro\second}-measurement time using a matched filter, as illustrated in Fig.~\ref{fig:MFA}(a,b). The peak ratio of bimodal Gaussian fits (with equal variance) to the readout-traces histograms of qubits initialized in the ground state (no pulse applied: \(\emptyset\)) provide insight in the thermal excitation probability \(\mathcal{P}(1\vert\emptyset)\). Comparing the peak ratios for trimodal Gaussian fits to the readout-traces histograms of qubits initialized in the excited state (\(\pi\)-pulse applied: \(\pi\)) indicate the conditional probability for qubit-energy decays \(\mathcal{P}(0\vert \pi)\) and second-excited state population \(\mathcal{P}(2\vert \pi)\). \(\mathcal{F}_{\rm label}=1-(\mathcal{P}(1\vert\emptyset)+\mathcal{P}(0\vert\pi))/2\) denotes a lower boundary for the initialization fidelity and thus the label accuracy using the conditional state transition rates. \(\mathcal{F}_{\rm \pi}\) represents the fitted \(\pi\)-pulse fidelities resulting in the preparation fidelities \(\mathcal{F}_{\rm prep}=(1+[1-2\mathcal{P}(1\vert\emptyset)]\mathcal{F}_{\rm \pi})/2\). \(\langle S_0\rangle\), \(\langle S_1\rangle\), and var(\(S\)) are the mean ground state, mean excited state, and variance of both states used to derive the Fisher criterion \(R\) and achievable assignment fidelity \(\mathcal{F}_{\rm ach}\) (see Eq.~\ref{eq:R},~\ref{eq:Fach}). \(\mathcal{F}^{\rm \overline{MF}}\), the product of \(\mathcal{F}_{\rm label}\) and \(\mathcal{F}_{\rm ach}\), is an estimate for an upper qubit-state-assignment fidelity bound for a classifier composed of a matched filter and the subsequent optimized threshold, here referred to as MF.}
\begin{tabular}{@{}c|cccc|ccc|cccc|c||c@{}}
\midrule\midrule
Qubit~~&~~\(\mathcal{P}(1_i|\emptyset_i)\)~~&~~\(\mathcal{P}(2_i|\emptyset_i)\)~~&~~\(\mathcal{P}(0_i|\pi_i)\)~~&~~\(\mathcal{P}(2_i|\pi_i)\)~~&~~\(\mathcal{F}_{\rm label}\)~&\(\mathcal{F}_{\rm \pi}\)&~\(\mathcal{F}_{\rm prep}\)~~~&~~~\(\langle S_0 \rangle\)~~&~~\(\langle S_1\rangle\)~~&~var(\(S\))~&\(R\)~~&~~\(\mathcal{F}_{\rm ach}\)~~&~~\(\mathcal{F}^{\rm \overline{MF}}\)~~\\\midrule
1       & 0.005 & \(\ll\)0.001    & 0.038 & ~~0.001       & ~0.979 & 0.999 & 0.995~     &~~1.061 & -0.947    & 0.388 & 26.817~~  & 0.995     & 0.974 \\
2       & 0.003 & \(\ll\)0.001    & 0.106 & ~~0.019       & ~0.946 & 0.977 & 0.986~     &~~0.523 & -1.145    & 0.963 & ~~3.001~~ & 0.807     & 0.773 \\
3       & 0.006 & \(\ll\)0.001    & 0.057 & ~~0.052       & ~0.968 & 0.965 & 0.977~     &~~0.731 & -1.181    & 0.355 & 28.927~~  & 0.996     & 0.965 \\
4       & 0.009 & ~~0.018         & 0.051 & ~~0.734       & ~0.961 & 0.970 & 0.976~     &~~1.003 & -0.101    & 0.247 & 19.953~~  & 0.987     & 0.950 \\
5       & 0.003 & \(\ll\)0.001    & 0.036 & \(\ll\)0.001  & ~0.981 & 0.976 & 0.985~     &~~0.852 & -1.164    & 0.348 & 33.614~~  & 0.998     & 0.979 \\
\midrule\midrule
\end{tabular}
\vspace{5mm}

	\centering
	\includegraphics[scale=0.5]{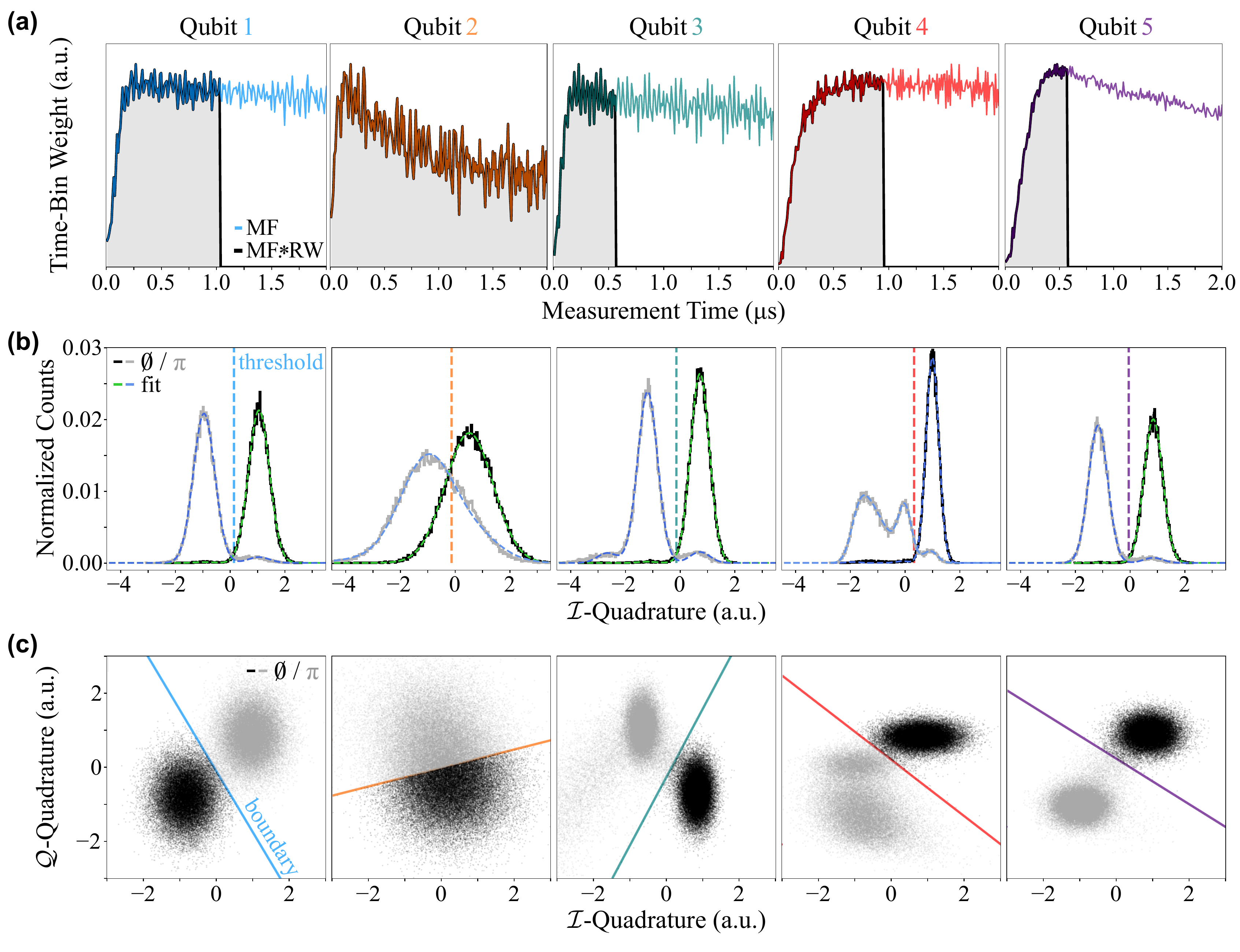}
    \captionof{figure}{\label{fig:MFA}Readout Data Statistics. (a) Magnitude of the time-bin weights of the qubit-specific matched filter shapes derived using prepared ground and excited states. A rectangular window (RW) is applied to each matched filter kernel to reduce the impact of qubit-energy decays and maximize qubit-state-assignment fidelities. The resulting matched filter windows are shaded in gray. (b) Shown are the histograms of the qubit-state-readout single-shot traces after applying the optimized \SI{1}{\micro\second}-long matched filter. The dashed lines represent the optimized thresholds with the states to the right attributed to the ground state and left to the excited state. Using bimodal Gaussian fit functions for the ground state (green) and trimodal Gaussian fit functions for the excited state (blue) provides insight into the underlying dynamics such as thermal excitation or qubit-energy decays (see Tab.~\ref{tab:MF}). (c) Plotted are boxcar filtered single-shot traces of ground (black) and excited states (gray) in the \(\mathcal{IQ}\)-plane. A linear support vector machine trained on the two-dimensional data generates the qubit-specific colored discrimination boundary.}
\end{figure*}

\subsection{\label{subsec:MF}Matched Filter (MF) Threshold Discriminator}
\ReMo{To reduce the computational discrimination effort, the elements of a measured single-shot readout trace are often summed up before a discriminator is applied. Filtering the readout traces before they are summed up further simplifies the discrimination process. Filtering in this context means multiplying each element \(n\), \( [S]_n = \mathcal{I}_n+j\mathcal{Q}_n \), of a discrete signal \(S=\left[[S]_1, [S]_2, \dots, [S]_N\right]\) by a window or kernel weight \(k_n\). If the weights are all unity over a particular range and zero otherwise, the filter is referred to as a boxcar filter. In general, the filtered signal \(\mathcal{S}\) can be computed as
\begin{equation}\label{eq_SumSig}
    \mathcal{S} = \sum^N_n k_n [S]_n = \sum_n k_n (\mathcal{I}_n+j\mathcal{Q}_n).
\end{equation}
For a boxcar filter, \(\mathcal{S}\) is a scalar complex number. The discrimination process is consequently a two-dimensional discrimination task. 

A matched filter, as we use the term in this paper, is a filter designed to optimize the signal-to-noise ratio (SNR), and projects the complex input signal to a single dimension. Hence, the resulting \(\mathcal{S}\) can be linearly separated~\cite{Turin1960_MF}. For two-class discrimination (such as in qubit readout) the matched filter is given by
\begin{equation}\label{eq_match}
    k_n=\frac{\langle [S_{0}]_n-[S_{1}]_n\rangle}{\text{var}([S_{0}]_n)+\text{var}([S_{1}]_n)},
\end{equation}
where \(S_{0}\) and \(S_{1}\) are the signals of the two classes (\(\langle f \rangle\) denotes the mean value of signal \(f\) and var(\(f\)) the variance of \(f\)). Assuming the noise in the signal is stationary and Gaussian distributed, this is the optimal weighting function~\cite{Bishop2006_Patt, Ryan2015_MF}, and the} optimized discriminator threshold is then located at \(0\), the axis origin~\cite{Ryan2015_MF}. 

For superconducting qubits, \ReMo{this matched filter} is equal to the difference between the mean ground- and excited-state-readout signals normalized by the signal variance, which must be measured experimentally using calibration runs with known qubit states---as described and termed ``matched filter'' in Ref.~\cite{Ryan2015_MF}, ``mode matched filter'' in Ref.~\cite{Heinsoo2018_fast_readout}, or as ``Fisher's linear discriminant'' in Ref.~\cite{Bishop2006_Patt}. \ReMo{While filtering is typically not considered as an example of a learning algorithm, the filter estimation} and threshold optimization \ReMo{can be thought of as} a ``training'' step. 

In our implementation, as illustrated in Fig.~\ref{fig:MFA}(a), the matched filter kernel is \ReMo{additionally} multiplied with a \ReMo{boxcar filter} to limit the impact of nonidealities such as qubit-energy decay. 
\ReMo{After matched filter summation (Eq.~\ref{eq_SumSig}), an optimized threshold} partitions the one-dimensional projection into ground- and excited-state classes, depicted in Fig.~\ref{fig:MFA}(b). Finally, the concatenation of the one-bit labels assigned by each single-qubit discriminator results in the assigned five-qubit-state label. Note, the demodulation step at intermediate frequencies using \(\text{e}^{\rm -j\omega^{IF}_in}\) with \(\omega_i^{\rm IF}\) defined in Tab.~\ref{tab:RES} (as described in Ref.~\cite{Krantz2019_review}) can be incorporated in the kernel tune-up.

\ReMo{Assuming the noise affects both qubit states equally}, the achievable assignment fidelity depends on the separation \(R\) between the ground- and excited-state-readout signals, \(S_0\) and \(S_1\), referred to as the Fisher criterion~\cite{Fisher1936_FC}. The separation \(R\) is defined as
\begin{equation}\label{eq:R}
    R=(\langle S_0\rangle-\langle S_1\rangle)^2/\text{var}(S), 
\end{equation}
\noindent
with \ReMo{the same} variance \ReMo{for both states}, \(\text{var}(S)=\text{var}(S_0)=\text{var}(S_1)\). For \ReMo{additive} Gaussian \ReMo{noise with a} diagonal covariance matrix, \(R\) \ReMo{is maximized by the} matched filter kernel of Eq.~\ref{eq_match}~\cite{Ryan2015_MF, Bishop2006_Patt}, 
with the maximally achievable assignment fidelity
\begin{equation}\label{eq:Fach}
    \mathcal{F}_{\rm ach} = \frac{1}{2}\left[1+\text{erf}\left(\sqrt{R/8}\right)\right],
\end{equation}
with \(\text{erf}(z)\), the Gauss error function of \(z\)~\ReMo{\cite{Magesan2015_SVM}}.

\ReMo{For each qubit state, the filtered-signal ($\mathcal{S})$} histograms that result after the matched filter are fit with Gaussian functions, shown in Fig.~\ref{fig:MFA}(b). For the fit functions, \ReMo{we assume the readout noise for both qubit states has the same variance, in order} to evaluate \ReMo{the maximally achievable discrimination fidelity} \(\mathcal{F}_{\rm ach}\) \ReMo{under ideal noise conditions}, as presented in Tab.~\ref{tab:MF}. Fitting the ground state with a bimodal, and the excited state with a trimodal Gaussian fit reveals \ReMo{nonidealities due to} state transition dynamics such as thermal excitations or qubit-energy decays. The product of the label, \(\mathcal{F}_{\rm label}\), and achievable, \(\mathcal{F}_{\rm ach}\), fidelities provides an estimation of the upper boundary for the matched filter (MF) discriminator qubit-state-assignment fidelity \(\mathcal{\mathcal{F}^{\rm \overline{MF}}}\), as shown in the last column of Tab.~\ref{tab:MF}.

\begin{figure}[t]
	\centering
	\includegraphics[width=0.45\textwidth]{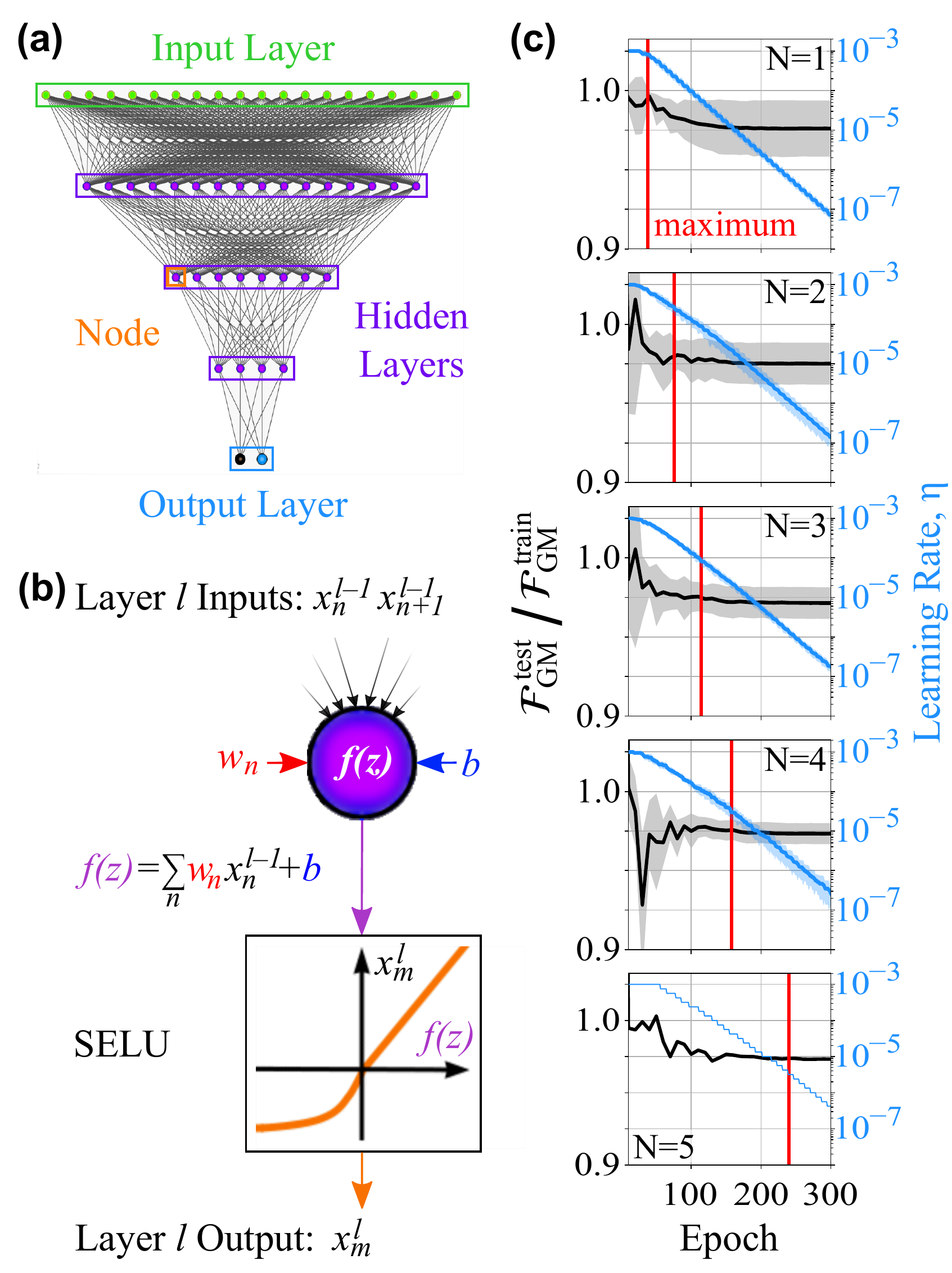}
	\caption{\label{fig:FNN}Architecture and Training of Fully-Connected Feedforward Neural Network (FNN). (a) The FNN architecture used here comprises an input layer, three hidden layers, and an output layer. For a \SI{1}{\micro\second}-long measurement time, the input layer consists of 1,000 nodes. 1,000, 500, and 250 nodes form the first, second, and third hidden layer. The output layer scales as \(2^{N}\) (N, the number of qubits). For five qubits, the output layer encompasses 32 nodes. (b) The nodes composing the hidden layer \(l\) are functions that depend on the following parameter inputs: the output values \(x_n^{l-1}\) of the prior layer \(l-1\) and a node-specific bias \(b\). The output value \(x_m^l\) of node \(m\) corresponds to the weighted (weights \(w_n\)) sum of the inputs \(x_n^{l-1}\) and the bias \(b\) after passing through an activation function, here a scaled exponential linear unit (SELU), shown in orange. (c) Shown is the training performance for an FNN tasked to discriminate \(N\) qubits with \(N={1,2,\dots,5}\). The generalization---the ratio of the geometric mean test \(\mathcal{F}_{\rm GM}^{\rm test}\) and training qubit-state-assignment fidelity \(\mathcal{F}_{\rm GM}^{\rm train}\)---as the number of epochs increases is shown in black using the left y-axis. The associated standard deviation of the generalization is indicated in gray. The number of epochs to achieve the maximum qubit-state-assignment fidelity is indicated with a red vertical bar. The learning rate \(\eta\), shown in blue and using the right y-axis, is gradually reduced as the number of epochs increases.}
\end{figure}

\subsection{\label{subsec:SVM}Support Vector Machine (SVM)}
Support vector machines (SVMs)---known for their robustness and good generalization---are fundamental two-class discriminators that draw a single decision boundary, called a hyperplane, in a supervised learning scheme~\cite{Boser1992_SVM, Cortes1995_SVM}. The margin between the classes and the hyperplane can be maximized by penalizing misclassified data points and data points within the margin boundaries. The penalty for data points within the margin boundaries can be varied using a regularization term. A lenient penalty results in a so-called soft-margin SVM which can better cope with problems that are not linearly-separable. 

The hyperplane dimension is equal to the one less than the number of features--the dimensions of the measurement data. The location of a new data point relative to the hyperplane decides on the associated label. This deterministic decision process is not probabilistic, and the information on the probability of label association is thus not directly accessible. While hyperplane separations only work for linearly-separable data, nonlinear SVMs use the kernel trick to map the data points to higher dimensions via a nonlinear transformation and find a hyperplane in that higher-order feature space. 

Several SVMs can be trained in concert for multi-class discrimination to divide the feature space into areas associated with distinct classes~\cite{DudaHart1973_MQSVM}. For an \textit{N}-class (\(N>2\)) classification task, the number of necessary hyperplanes is at least \(N-1\) if each class is discriminated against the rest, referred to as ``one-versus-all.'' Each class requires a hyperplane separating itself from the remaining collective of classes. However, separating space in more than two classes results in ambiguous areas that cannot be associated with a single class~\cite{Bishop2006_Patt}.  

Here, we use scikit-learn library to implement single-qubit and multi-qubit linear and nonlinear SVMs in Python~\cite{buitinck2013_Sklearn}. We employ the LinearSVC implementation for linear and SVC for nonlinear soft-margin SVMs with regularization parameters optimized per discriminator to deliver the maximally achievable qubit-state-assignment fidelity. In general, the training wall-clock-time for an SVM implemented using LinearSVC is significantly reduced relative to the training time required for SVC SVMs. Nonlinear SVMs can only be implemented in SVC, as LinearSVC does not offer the kernel trick. In addition to the resulting unfavorable scaling of the training wall-clock-time of nonlinear SVMs, the multi-dimensional optimization problem, if tasked to discriminate multiple qubit states, mostly resulted in non-optimal hyperplanes (for five qubits, nonlinear SVMs achieved an average qubit-state-assignment fidelity about \(\SI{10}{\percent}\) worse than the one achieved by its linear counterpart). We limit the study of nonlinear SVMs to a basic characterization due to the lack of qubit-state-assignment fidelity robustness and the training-time requirements (for five qubits more than one day). Henceforth, we focus on linear soft-margin SVMs as parallel single-qubit or multi-qubit discriminators (in the one-versus-all mode). 

\begin{figure}[t]
	\centering
	\includegraphics[width=0.45\textwidth]{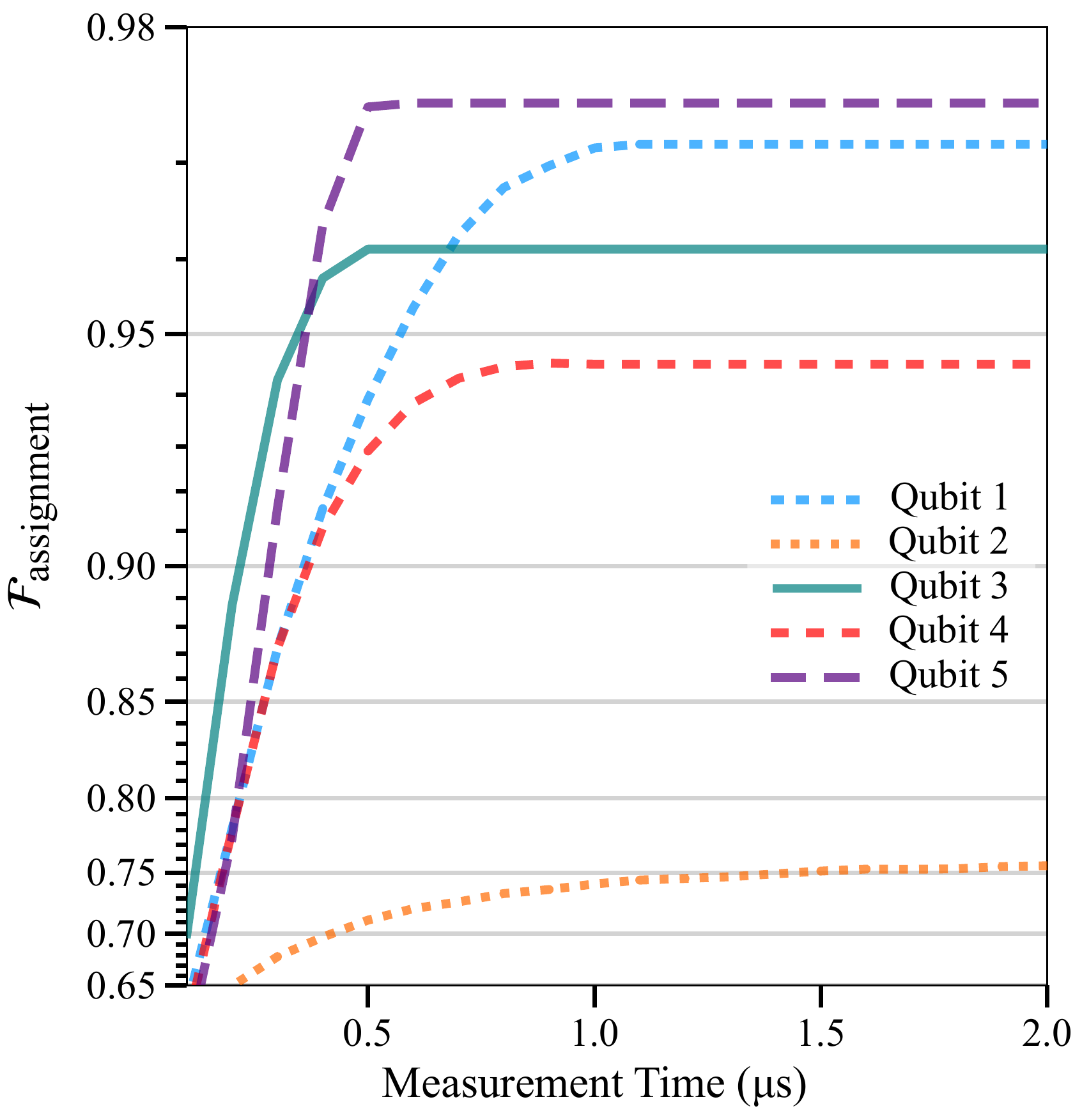}
	\caption{\label{fig:MFT}Qubit-State-Assignment Fidelity. Matched filter discriminator for each qubit versus measurement time. The maximum assignment fidelity \(\mathcal{F}_{i}(t_i)\) for each qubit \(i\) is reached after \(t_{\color{Q1}{1}}=\SI{1}{\micro\second}\), \(t_{\color{Q2}{2}}=\SI{2}{\micro\second}\), \(t_{\color{Q3}{3}}=\SI{0.5}{\micro\second}\), \(t_{\color{Q4}{4}}=\SI{0.8}{\micro\second}\), and \(t_{\color{Q5}{5}}=\SI{0.5}{\micro\second}\).}
\end{figure}

\begin{figure*}[t]
	\centering
	\includegraphics[width=0.94\textwidth]{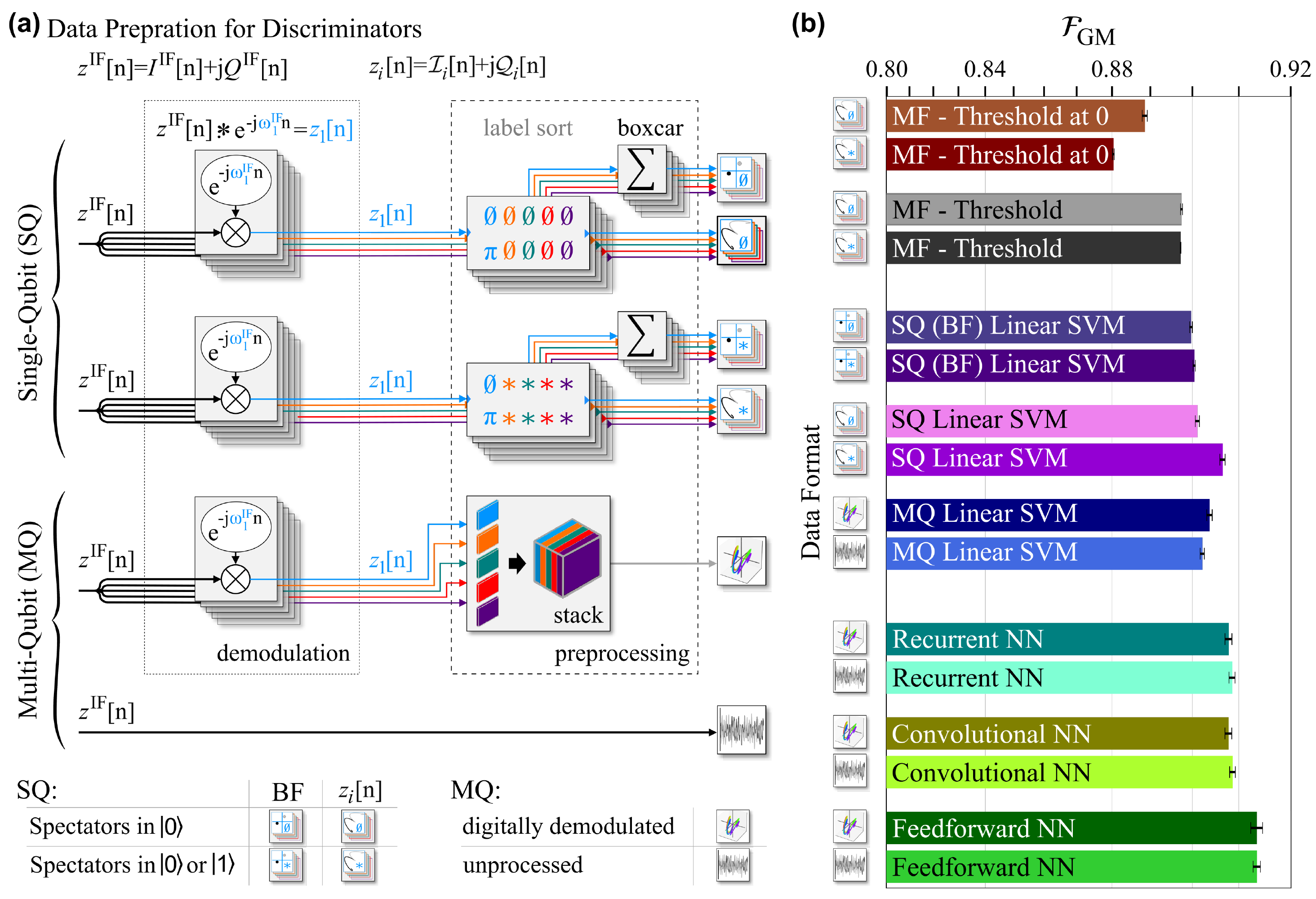}
    \caption{\label{fig:DCO}Measurement Data Processing and Discrimination. (a) \(M\)-dimensional data (\(z^{\rm IF}\)[n]) processing for single-qubit (SQ) and multi-qubit (MQ) discrimination. For single-qubit discrimination, \(z^{\rm IF}\)[n] is digitally demodulated at the intermediate frequency of a resonator \(i\). The resulting signal \(z_{i}\)[n] can be simplified with a boxcar filter (BF) [\(\frac{1}{M}\sum_n z_{i}\text{[n]}=\bar{\mathcal{I}}_i+j\bar{\mathcal{Q}}\)]  or kept as sequences \(\mathcal{I}_i\)[n] and \(\mathcal{Q}_i\)[n]. The discriminators can either be trained with the spectator qubits exclusively in their ground state (denoted by \(\emptyset\)) or, alternatively, in either their ground or excited state (denoted by \(\ast\)). For multi-qubit discriminators, the digitally demodulated signals \(z_{i}\)[n] at all resonator frequencies \(i\) are stacked up. The resulting data block is subsequently used for the discriminator training. Alternatively, the discriminator can be tasked to discriminate \(z^{\rm IF}\)[n] directly without any digital preprocessing. (b) Comparison of the geometric mean qubit-state-assignment fidelity for five qubits after a \SI{1}{\micro\second}-long measurement and 10,000 training instances per qubit-state configuration. All single-qubit discriminators are evaluated using training data with the spectator qubits in the ground as well as all combinations of ground and excited state. The matched filter (MF) threshold discriminator [the matched filter is part of the discriminator and thus not shown in (a)] is shown in two configurations; the threshold set to \(0\) and the threshold optimized. The linear support vector machine (SVM) is applied to boxcar-filtered (BF) and time-trace data of \(\mathcal{I}_i\)[n] and \(\mathcal{Q}_i\)[n]. The multi-qubit discriminators are evaluated utilizing digitally demodulated and unprocessed data. Shown are a multi-qubit linear SVM, a recurrent neural network (NN), a convolutional NN, and feedforward NN.}
\end{figure*}

\subsection{\label{subsec:NN}Neural Networks (NN)}
Typically, a neural network consists of an input layer composed of several nodes---the number of nodes depends on the input data dimension---and an output layer that contains the computed output values. In between the input and output layer are layers of neurons--- so-called hidden layers as their output value is not directly accessible---with unique tasks per layer. The input and output channels of a neuron are called edges, illustrated in Fig.~\ref{fig:FNN}(a). Each neuron can be described as a mathematical function of incoming weighted parameters---typically output values of other neurons---and external parameters. The function output generally passes through a nonlinear filter before it can serve as an input to other neurons, depicted in Fig.~\ref{fig:FNN}(b). Varying the connectivity, neuron functions, and the nonlinear function at each neuron output provides a flexible toolset to engineer a broad spectrum of neural network types. Supervised training of such a network can optimize the weights for each neuron input and external parameter to almost arbitrarily approximate any function. 

\begin{figure*}[t]
	\centering
	\includegraphics[width=0.94\textwidth]{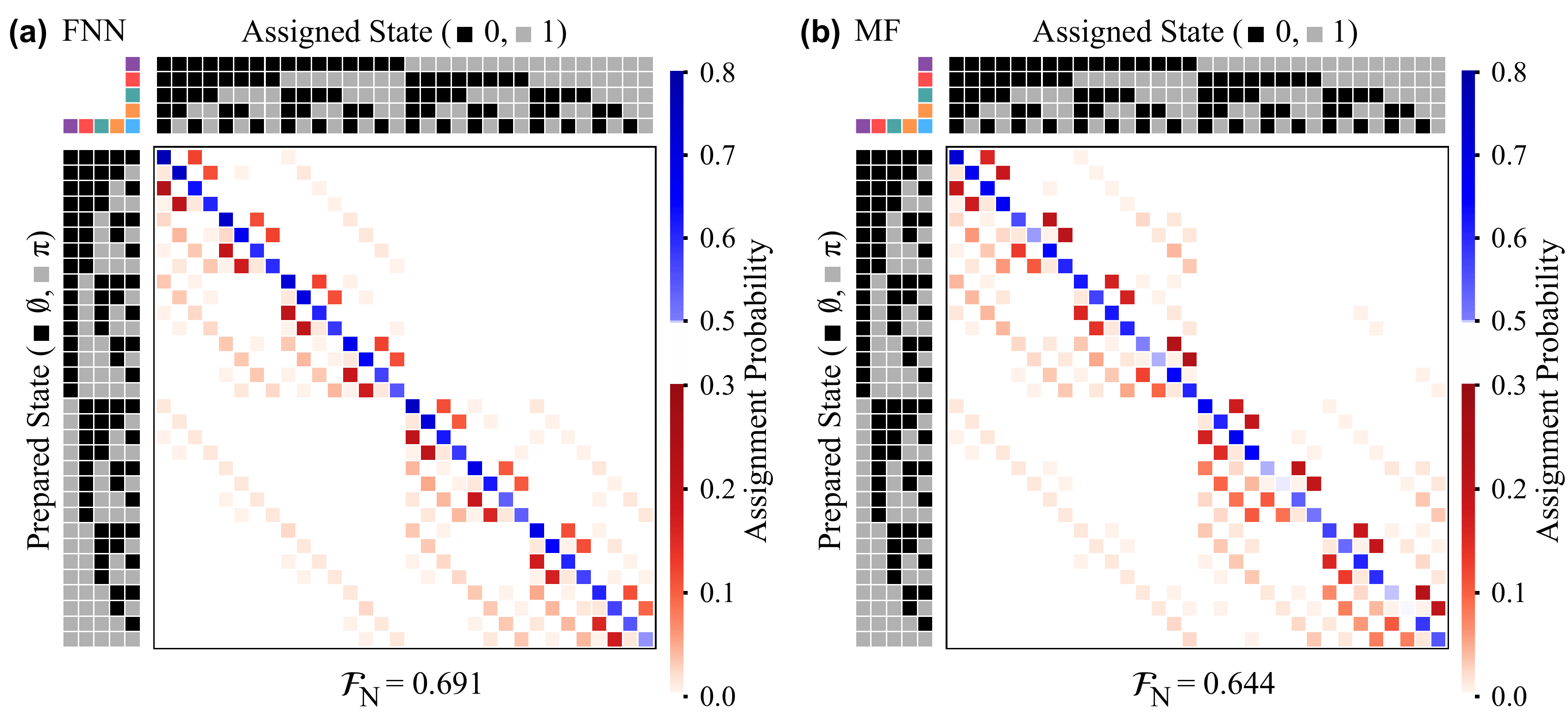}
    \caption{\label{fig:CON}Qubit-State-Assignment Fidelity Analysis. Confusion (assignment probability) matrix of the feedforward neural network (FNN) (a) and matched filter (MF) (b). The rows of the confusion matrix encompass the probability distribution of the discriminator to assign each of the 32 qubit-state configurations to the row's prepared qubit-state configuration (no pulse applied, qubit initialized in the ground state: \(\emptyset\rightarrow 0\); \(\pi\)-pulse applied, qubit initialized in the excited state: \(\pi\rightarrow 1\)). The probabilities of correctly classified states--on the diagonal--are shown in blue, whereas the misclassification probabilities--the off-diagonals--are shown in red. \(\mathcal{F}_{\rm N}\), introduced in Eq.~\ref{eq:FFN}, represents a metric to indicate the overlap between the confusion matrix and an identity matrix (the ideal confusion matrix). \(\mathcal{F}_{\rm N}=1\) if the confusion matrix is an identity matrix.}
\end{figure*}

We have examined various neural network architectures to determine the most useful one in improving the qubit-state assignment fidelity and measurement time of multi-qubit devices. We have explored fully-connected feedforward neural networks (FNN)---among the most elementary neural networks---convolutional neural networks (CNN)---among the most successful image classification methods in use today---and long short-term memory recurrent neural networks (LSTM)---among the most successful architectures in language processing. The fully-connected FNN with three hidden layers excelled in assignment fidelity compared to the other neural network types. 

Implemented in PyTorch~\cite{Paszke2019_pytorch}, the FNN architecture that yields the highest assignment fidelity for five qubits is composed of three hidden layers. The number of nodes composing the input layer depends on the measurement time and the size of the discrete time-bins---here \SI{2}{\nano\second}. For a \SI{1}{\micro\second}-long measurement time, the input layer contains 1,000 nodes with the in-phase and quadrature components alternating. The dimension of the first hidden layer is equal to, the second hidden layer is half of, and the third hidden layer is a quarter of the input layer dimension. Finally, the output layer consists of \(2^{N}\) nodes, with \textit{N} being the number of qubits (32 for the five-qubit readout we focus on here). The activation function, the nonlinear filter acting on the hidden layer nodes, is a scaled exponential linear unit (SELU)~\cite{Klambauer2017_SELU}, instead of the common rectified linear unit (ReLU)~\cite{Nair2010_ReLU} due to its improved robustness and learning rate. The output layer is filtered using a softmax function \(\text{softmax}(x_i)=\exp(x_i)/\sum_j{\exp(x_j)}\). 

\ReMo{The architectural complexity of the neural network architecture depends on the number of time bins constituting each measurement, the number of multiplexed frequencies, and the number of qubits. For our investigation, we found that the FNN requires at least two and optimally three layers. The first hidden layer has the length of the input layer. The consecutive layers should then have half the number of nodes of the previous layer. While we did not observe an improvement in adding more nodes to the layers, we observed a decrease in assignment fidelity when the layers comprise fewer than half the nodes of the prior layer.}

\ReMo{It may be possible to reduce the complexity of the neural network if the number of available training samples is limited. We found that for a training set of 100 samples per qubit state, a feedforward neural network consisting of a single hidden layer and 10 nodes is sufficient for the readout of the superconducting qubit system described here~\cite{Riste2020_RealTimeProcessing}. For 20 randomized training sets of 100 samples per state, the matched filter reached an assignment fidelity of \(58.9\% \pm 3.4\%\), whereas the feedforward neural network yielded \(80.0\% \pm 2.7\%\). For 5100 samples per state, the assignment fidelity was comparable for both discriminators: 88.8\% for the matched filter and 89.4\% for the feedforward neural network. In general, for small training sets, the distribution of rare effects such as excited state decays is not well balanced and thus a training bias is to be expected. The considerable error bar is a consequence of that training bias. Therefore, larger training sets are typically preferred.}


Multiple training cycles, referred to as epochs, are required to ensure the discriminator output to converge to the maximum qubit-state-assignment fidelity. The number of epochs to reach a convergence plateau depends on the correction factor per cycle, the learning rate. We start with a more aggressive learning rate of \(0.001\)---a typical value for neural networks---and gradually decrease it as the qubit-state-assignment fidelity starts plateauing around \(250\) epochs. Furthermore, the entire training set is randomly divided into normalized sub training units, termed batches~\cite{Bengio2012_TNN}. The batch size specifies after how many training samples the neural network weights are updated. The choice of batch size affects the wall-clock-training time and generalization, or in other words, how well the discriminator performs on unseen data compared to the training set. We find that a batch size of 1,024 achieves a good balance between assignment fidelity, generalization, and wall-clock-training time. We observe an average wall-clock-training time of about half an hour for five qubits. The learning rate, generalization, and the optimal number of epochs as the number of qubits increases is shown in Fig.~\ref{fig:FNN}(c). 

\section{\label{sec:REA}Result Analysis}

In addition to a specific choice of discriminator, the to-be-discriminated data can be differently prepared. Typically, the discrete time readout signals at intermediate frequency, \(z^{\rm IF}\text{[n]}=I^{\rm IF}\text{[n]}+jQ^{\rm IF}\text{[n]}\), are digitally demodulated following the steps outlined in Fig.~\ref{fig:DCO}(a) and Ref~\cite{Krantz2019_review}. The signal components \(\mathcal{I}_i\text{[n]}=\Re\left(z_{i}\text{[n]}\right)\) and \(\mathcal{Q}_i\text{[n]}=\Im\left(z_{i}\text{[n]}\right)\) can be boxcar filtered~\cite{Krantz2019_review} or kept as a sequences \(\mathcal{I}_i\text{[n]}\) and \(\mathcal{Q}_i\text{[n]}\). For digitally demodulated data and multi-qubit discrimination, \(z^{\rm IF}\)[n] are demodulated at each intermediate frequency. The resulting digitally demodulated time traces need to be stacked up to form a single data block before used as the input to the multi-qubit discriminator. 

Furthermore, the training data set can be either composed of all permutations of the qubit states or a specific subset. Here, we focus on either training discriminators with qubits not involved in the training process, the spectator qubits, in all combinations of the ground and excited state (indicated as \(\ast\)), or kept in the ground state (denoted by \(\emptyset\)). 

We evaluate the comparison for a measurement time of \SI{1}{\micro\second} after which four out of five qubits have reached their maximum assignment fidelity for matched filters, as shown in Fig.~\ref{fig:MFT}. For five qubits, a \SI{1}{\micro\second}-long measurement time, and 10,000 training instances, we show a comparison of the qubit-state-assignment fidelity of the above introduced single- and multi-qubit discriminator approaches in Fig.~\ref{fig:DCO}(b). Optimizing the threshold of MFs and using training data with the spectator qubits in the ground state increases the qubit-state-assignment fidelity. Single-qubit linear SVMs perform best if tasked to discriminate vectorized digitally-demodulated data and trained with a data set with all qubit-state combinations represented.  

Multi-qubit linear SVMs appear to perform better if tasked to discriminate digitally demodulated readout signals. On the contrary, the neural networks perform the best if unprocessed data is used. The feedforward neural network outperforms its counterparts, the recurrent and convolutional neural network, in the achieved qubit-state-assignment fidelity. \ReMo{The RNN processes the data chronologically, whereas the CNN performs temporally local operations. The fully-connected layers of the FNN process data without the notion of time. We suspect that the FNN outperforms its neural network archetype alternatives due to its temporally unbiased approach and robust training routine.}

In the main part of the manuscript, we focus on the best performing discriminator approach of each category: matched filter, single-qubit linear SVM, multi-qubit linear SVM, and neural networks. 

Next, we analyze the qubit-state-assignment probabilities using the metric of confusion matrices. Fig.~\ref{fig:CON} illustrates the confusion matrix for the FNN and MF discriminator. For an ideal confusion matrix with all prepared states agreeing with the assigned state, the confusion matrix is an identity matrix. To evaluate the overlap between an identity matrix (entries represented as a Kronecker delta \(\delta_{ij}\) with i and j representing the indices of the matrix row and column) and a confusion matrix (with entries \(c_{ij}\)), we propose the following metric based on the Frobenius norm 
\begin{equation}\label{eq:FRO}
    \vert\vert \text{A} \vert\vert_{\rm F} = \sqrt{\sum_i\sum_j\vert c_{ij}-\delta_{ij}\vert^2}.
\end{equation}
\\
To bound the Frobenius norm between 1 and 0, we normalize the Frobenius norm with the maximum value of Eq.~\ref{eq:FRO} (\(\sqrt{2^{N+1}}\)). The normalized Frobenius norm is equal to 0 if the confusion matrix is exactly an identity matrix. An alternative representation more closely related to the fidelity metric can be expressed as
\begin{equation}\label{eq:FFN}
    \mathcal{F}_{\rm N}=1-\frac{\vert\vert \text{A} \vert\vert_{\rm F}}{\sqrt{2^{N+1}}}.
\end{equation}
The MF achieves \(\mathcal{F}_{\rm N}=0.644\), whereas the FNN yields a value of \(\mathcal{F}_{\rm N}=0.691\), a relative improvement of \SI{7.3}{\percent}.
\bibliography{main}
\end{document}